\documentclass{jaa}

\usepackage{lmodern}
\usepackage{xcolor}
\usepackage{graphicx}
\usepackage{epstopdf}
\usepackage{lastpage}
\usepackage{multicol}
\usepackage{multirow}
\usepackage{array}
\usepackage[normalem]{ulem}
\usepackage{amssymb}
\usepackage{amsmath}
\usepackage{booktabs} 
\usepackage{xspace}
\usepackage{adjustbox}
\usepackage{lineno}
\usepackage{setspace}

\usepackage{subcaption}
\definecolor{xlinkcolor}{cmyk}{1,1,0,0}
\usepackage{natbib}
 \usepackage[bookmarks=false,         
     pdfnewwindow=true,      
     colorlinks=true,    
     linkcolor=xlinkcolor,     
     citecolor=xlinkcolor,     
     filecolor=xlinkcolor,  
     urlcolor=xlinkcolor,      
final=true 
 ]{hyperref}

\newcommand{\asat}{\textit{AstroSat}\xspace}
\newcommand{\tnbs}{thermonuclear bursts\xspace}
\newcommand{\source}{4U\:1728--34\xspace}

\newcommand{\chisq}{$\chi^2$\xspace}
\newcommand{\countsps}{counts~s$^{-1}$\xspace}

\begin{document}
\sloppy

\title{``Bursts, Beats, and Beyond": Uncovering the landscape from accretion to ignition of \source using \textit{AstroSat}}



\author{\href{https://orcid.org/0000-0003-3173-4691}{Anirudh Salgundi}\textsuperscript{1}, 
\href{https://orcid.org/0000-0002-6657-9022}{Suman Bala}\textsuperscript{2,1}, 
\href{https://orcid.org/0000-0003-0852-3685}{Gayathri Raman}\textsuperscript{3},
\href{https://orcid.org/0009-0002-7897-6110}{Utkarsh Pathak}\textsuperscript{1},
\href{https://orcid.org/0000-0002-6112-7609}{Varun Bhalerao}\textsuperscript{1}}

\affilOne{\textsuperscript{1} \footnotesize Department of Physics, Indian Institute of Technology Bombay, Powai, Mumbai 400076, India.\\}
\affilTwo{\textsuperscript{2} Science and Technology Institute, Universities Space Research Association, Huntsville, AL 35805, USA.\\}
\affilThree{\textsuperscript{3} Department of Astronomy and Astrophysics, The Pennsylvania State University, 525 Davey Lab, University Park, PA 16802, USA.\\}


\twocolumn[{

\maketitle

\corres{anirudhsalgundi@gmail.com}


\sloppy
\begin{abstract}
A comprehensive study on persistent and thermonuclear burst emission of \source, commonly known as `Slow Burster' is performed using seven archival observations of \textit{AstroSat} spanning from 2016--2019. The burst-free persistent spectra can be well fitted with a blackbody \texttt{bbody} and a powerlaw \texttt{powerlaw} components, with a powerlaw photon index ($\Gamma$) was found to be $\sim$2 indicating the source was in ``high/soft" bananna state or intermediate state. The time averaged power density spectrum reveals the presence of twin kilohertz Quasi Periodic Oscillations (kHz QPOs) with centroid frequencies $619\pm10$~Hz and $965\pm6$~Hz with a maximum fractional root mean squared amplitude of $6.24\pm1.31$~\% at $\sim$16~keV. From the upper kHz QPO, we infer the magnetospheric disk radius to be $\sim$17~km, corresponding to a magnetic field strength of 0.35--1.27$~\times~10^7$~G. The burst spectral evolution indicates Photospheric Radius Expansion (PRE) in five bursts, yeilding a touchdown radius of 3.1--5.47~km. These bursts reached near-Eddington luminosities, through which the distance of the source was calculated to be 5.18--5.21~kpc. Two of the bursts show coherent oscillations at 362.81--363.93~Hz. The presence of twin kHz QPOs and coherent Burst Oscillations allows us to provide two different estimates for the spin frequency of the Neutron Star in the system, for the first time using \asat.

\end{abstract}

}]

\section{Introduction}
 
Low-Mass X-ray Binaries (LMXBs) are a class of binary systems that harbour a compact object either a Neutron Star (NS) or a Black Hole (BH) that accretes matter from a low mass ($\lesssim$1~M$_\odot$) companion star. Accretion in these systems typically proceeds through Roche Lobe overflow. Depending on their spectral evolution properties, these systems are further classified into Atoll and Z sources \citep{Hasinger1989AA}. Atoll sources, in particular, transitions between what are known as low intensity `island' state (or hard state) and high intensity `banana' state (or soft state) based on the shape they trace on the colour-colour diagram. These state transitions have been associated with changes in mass accretion rate and evolution of the accretion flow \citep{Hasinger1989AA}. Unlike the Z class of sources, the Atoll source achieves maximum luminosities of $\le10\%~L_\mathrm{Edd}$, varying less dynamically along the spectral state branches. The Neutron Stars (NS) in these systems are found to have low surface magnetic fields ($\sim$10$^8$~G). 

Atoll sources are known to exhibit several types of time variability phenomena such as Quasi Periodic Oscillations (QPOs) and Thermonuclear Bursts (TNBs). QPOs are coherent signals observed in the Power Density Spectra, typically in mHz to kHz range. There is no consensus regarding the origin of the QPOs. They may originate due to the interaction between magnetosphere and the accretion flow at the innermost region of the accretion disk or due to relativistic effects near the NS surface \citep{Wang2016}. For several sources, the frequencies and amplitudes of these features are found to be correlated with the luminosity states \citep{Ford2000ApJ}, position in the Hardness Intensity Diagram (HID) \citep{Wijnands1997ApJ}, and even the spectral power law indices \citep{Kaaret2002ApJ}. Almost all Atoll sources also exhibit \tnbs. TNBs occur due to unstable nuclear burning of accreted material (predominantly composed of Hydrogen and Helium) on the surface of the NS \citep{Bhattacharyya2010AdSpR, Galloway2003ApJ}. The luminosities during these burst episodes can reach Eddington levels in a few seconds. The X-ray burst spectra of TNBs are typically characterized using blackbody emission. A small fraction ($\le$20~\%) of these bursts exhibit rotational modulations of temperature variations, which are called `Burst Oscillations' (BO). BOs are coherent, periodic variations in the burst flux. Their peak frequencies are associated with the NS spin frequency \citep{Wijnands1998Nature}, thus making TNBs an important tool for probing NS spin.


The source, \source (commonly known as the `slow burster') is a persistent NS-LMXB that has been extensively characterized for over 5 decades using a number of X-ray telescopes. The NS nature of the compact object in this source was established soon after the detection of the first set of \tnbs~using \textit{SAS-3} and \textit{Uhuru} \citep{Hoffman1976ApJ, Kellogg1971ApJ, Lwein1976IAUC, Basinska1984ApJ}. This LMXB is now known to host a weakly magnetized NS that is accreting from a Hydrogen poor companion \citep{Shaposhnikov2003ApJ, Vincentelli2023MNRAS}. Despite being a persistent Atoll source with no dramatic luminosity fluctuations, \source exhibits a wide range of movement in its HID, where it transitions between soft and hard spectral states every $\sim$40--60~days \citep{MunozDarias2014MNRAS, Kong1998NewA}.

\source exhibits regular \tnbs which have been detected and characterized using \textit{RXTE} \citep{Strohmayer1996ApJ, Shaposhnikov2003ApJ}, \textit{XMM}-NEWTON \citep{Wang2019MNRAS}, \textit{INTEGRAL} \citep{Falanga2006AA} and more recently, using \textit{NICER} \citep{Mahmoodifar2019ApJ} and \asat \citep{VerdhanChauhan2017ApJ} as well. There are more than 1100 bursts reported for this source (see the Multi Instrument Burst ARchive (MINBAR)\footnote{\url{https://burst.sci.monash.edu/minbar/}} catalogue for most updated numbers;  \citep{Galloway2020}). The source distance, constrained using Eddington-limited Photospheric Radius Expansion (PRE) bursts, is around 4.4--5.1~kpc  \citep{DiSalvo2000ApJ, Galloway2003ApJ}. The \tnbs from the \source are usually short duration helium dominated bursts with typical burst recurrence
time of $\sim$3~hours \citep{Vincentelli2020MNRAS}. Burst Oscillations (BO) at a frequency of $\sim$361--363~Hz have been reported for this source. The BO frequencies were typically found to increase as the burst evolved \citep{vanStraaten2001ApJ, VerdhanChauhan2017ApJ, Franco2001ApJ}. A recent study of the BO phenomenon using \textit{NICER} reported an unusually strong set of oscillations that had widely different properties in different energy bands \citep{Mahmoodifar2019ApJ}. This was the first bursting source where an infrared counterpart to a TNB was observed. The delay between the X-ray and IR burst emission has been used to constrain the orbital period of the system to be $\le~$3~hours at an inclination angle $\ge~$8$^{\circ}$ \citep{Vincentelli2020MNRAS, Vincentelli2023MNRAS}. 

\source exhibits strong aperiodic variability and millisecond-timescale Quasi-Periodic Oscillations (kHz QPOs), including twin kHz QPOs \citep{DiSalvo2001ApJ,Mendez2001ApJ}. The intriguing `parallel track' behaviour between the lower kHz QPO and the X-ray intensity has been ascribed to bimodal accretion flows with contributions from disc and radial accretion \citep{WangZhang2020MNRAS,Mendez2001ApJ}.

The broadband X-ray spectrum for this source is described using a thermal comptonization model along with a blackbody component (for example, using \textit{INTEGRAL}: \citep{Falanga2006AA}, \textit{BeppoSAX}: \citep{DiSalvo2000ApJ}). Additionally, the source has been observed to exhibit a strong reflection component \citep{Mondal2017MNRAS, Wang2019MNRAS} in addition to a broad Fe line (see, for example, \citep{DiSalvo2000ApJ, Tarana2011MNRAS}). 


\asat-LAXPC stands out as one of the few currently operational X-ray instruments, with high timing capabilities and broad spectral coverage, making it an ideal tool for investigating bursting sources and their properties. In this paper, we present a comprehensive analysis of several \asat observations of \source and characterize the properties of the detected TNBs as well as the persistent emission. 

The paper is organized as follows: In Section \ref{sec:datareduction}, we describe the observations and data reduction methods adopted. We present the data analysis methods and results in Section \ref{sec:dataanalysis}. We summarize our results and discuss our findings in Section \ref{sec:discussion}.

\section{Observations and Data reduction}\label{sec:datareduction}
 
\asat is India's first space-based multi-wavelength observatory, which is designed to perform simultaneous multi-wavelength observations ranging from far UV to hard X-rays. It consists of five sets of instruments, namely Ultraviolet Imaging Telescope (UVIT) \citep{tandon_uvit}, Soft X-ray Telescope (SXT) \citep{SXT_singh_et_al}, Large Area X-ray Proportional Counter (LAXPC) \citep{2017AntiaLAXPC}, Cadmium Zinc Telluride Imager (CZTI) \citep{Bhalerao2017JApA} and Scanning Sky Monitor (SSM) \citep{Ramadevi_SSM} onboard the satellite.

\begin{table*}
    \centering
    \renewcommand{\arraystretch}{1.2}
    \scriptsize
    \caption{Observation details of \source.}
    \label{tab:obsdetails}
    \begin{adjustbox}{width=\textwidth,center}
    \begin{tabular}{|c|c|c|c|c|c|c|}
        \hline
        \textbf{Observation ID} & \textbf{Start time} & \multicolumn{2}{c|}{\textbf{Exposure (ks)}} & \multicolumn{2}{c|}{\textbf{Mean count rates}} & \textbf{Number} \\
        \cline{3-6}
         & \textbf{(MJD)} & \textbf{SXT} & \textbf{LAXPC} & \textbf{SXT} & \textbf{LAXPC} &  \textbf{of bursts}\\
         &  & &  & \textbf{(0.6--7.0~keV)} & \textbf{(3.0--80.0~keV)} & \\
        \hline
        9000000578 (O1) & 57606.36 & 13.26 & 64.70 & $9.42 \pm 3.07$ & $472.02 \pm 21.73 ^{L1,2,3}$ & 1 \\
        9000001904 (O2) & 58168.78 & 3.84 & 21.22 & $3.47 \pm 1.86$ & $478.57 \pm 21.88 ^{L2}$ & 0 \\
        9000002234 (O3) & 58316.75 & 7.58 & 21.07 & $3.52 \pm 1.88$ & $383.39 \pm 19.58 ^{L2}$ & 0 \\
        9000002254 (O4)$^a$ & 58324.11 & -- & 108.20 & -- & $261.38 \pm 16.17 ^{L2}$ & 2 \\
        9000002268 (O5) & 58331.98 & 4.36 & 23.88 & $5.18 \pm 2.28$ & $281.95 \pm 16.79 ^{L2}$ & 0 \\
        9000002890 (O6) & 58607.47 & 56.58 & 206.00 & $9.26 \pm 3.04$ & $274.61 \pm 16.57 ^{L2}$ & 5 \\
        9000003134 (O7) & 58726.23 & 38.13 & 187.34 & $9.42 \pm 3.07$ & $353.40 \pm 18.80 ^{L2}$ & 5 \\
        \hline
    \end{tabular}
    \end{adjustbox}
    \vspace{2mm}
    \footnotesize{\textsuperscript{a} SXT was not pointing at the source during O4, $^{L1,2,3}$ All three LAXPC detectors were used, $^{L2}$ only LAXPC20 was used.}
\end{table*}

\begin{figure}
    \centering
    \includegraphics[width=\columnwidth, height=6cm]{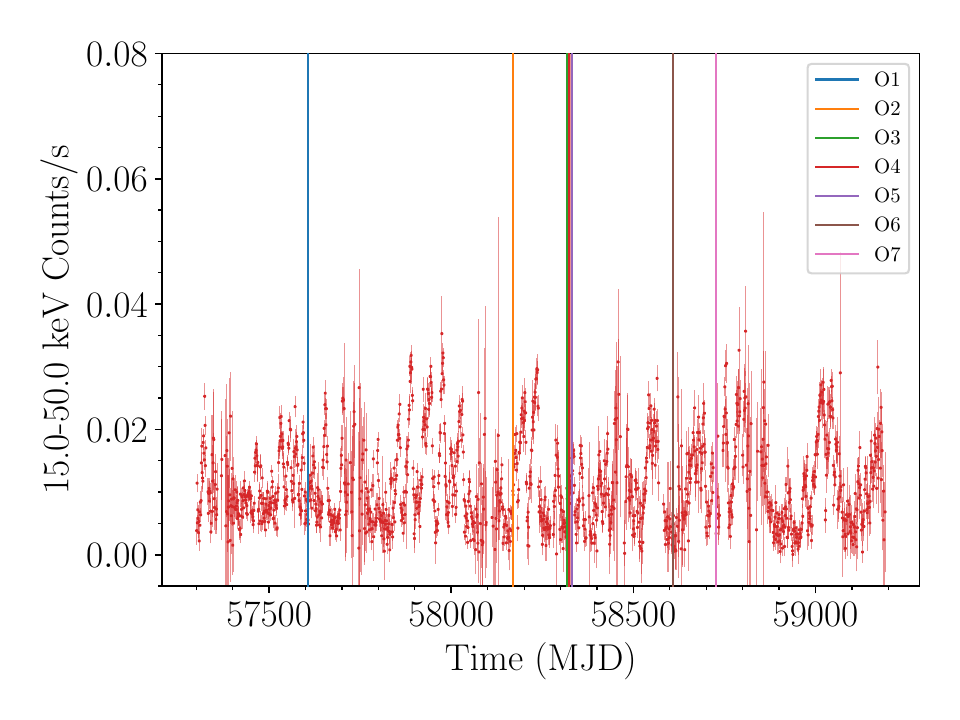}
    \caption{\textit{Swift}-BAT daily lightcurve\protect\footnotemark{} of \source in the 15.0-50.0~keV energy range. The vertical line indicates epochs of \asat observations (O1--O7, in order)used for this work.}
    \label{fig:swift_lc}
\end{figure}

For this study, we have utilized the archival observations of \asat\footnote{\url{https://astrobrowse.issdc.gov.in/astro_archive/archive/Home.jsp}} on \source taken using LAXPC and SXT instruments between 2016--2019 (see Figure \ref{fig:swift_lc}). A total of 13 \tnbs were detected in these observations. Further details regarding the observations are presented in Table~\ref{tab:obsdetails}.

\footnotetext{\url{https://swift.gsfc.nasa.gov/results/transients/GX354-0/}}

\subsection{SXT}
SXT is an imaging and focusing telescope operating in soft X-rays between 0.3--8.0~keV. It has an effective area of approximately 90~cm$^2$ at $\sim$1.5~keV, with a temporal resolution of $\sim$2.4~s and a spectral resolution of $\sim$150~eV at 6~keV in the Photon Counting (PC) mode. The level 1 SXT data were processed using the standard reduction pipeline \texttt{AS1SXTLevel2-1.4b}\footnote{\url{http://astrosat-ssc.iucaa.in/sxtData}} for each of the orbits and finally merged them using SXT merger tool \texttt{SXTPYJULIAMERGER v01}. The source image was extracted from the merged event file using \texttt{XSELECT V2.5b} from which the lightcurves and spectra were extracted. The background and response matrix files are provided by SXT Payload Operations Centre team. An off-axis Ancillary Response Files (ARF) was generated using the \texttt{sxt\_ARFModule} tool wherever necessary.

\subsection{LAXPC}

LAXPC is a cluster of three co-aligned proportional counters LAXPC-10, LAXPC-20 and LAXPC-30, that operates between 3.0--80.0~keV energy range with a total effective area of 6000~cm$^2$. It has an absolute temporal resolution of 10~$\mu$s and a deadtime of 42~$\mu$s \citep{antia2017calibration}. The level 1 LAXPC data were processed and background subtracted lightcurves and spectra were obtained using the standard data analysis tools  \texttt{LAXPCsoftware22Aug15}, which is made available on \asat Science Support Cell (ASSC).\footnote{\url{http://astrosat-ssc.iucaa.in/laxpcData}} Due to the gain instability caused by the gas leakage in LAXPC-10 since March 28, 2018 and LAXPC-30 not operational during some observations, data from LAXPC-20 was used to carry out spectral and temporal analysis for all the observations along with LAXPC-10 and LAXPC-30 for the observations where they were operational.

\section{Data analysis and results}\label{sec:dataanalysis}
 
\subsection{Persistent emission, Lightcurves and Hardness Intensity Diagram}
\sloppy
The Good Time Intervals (GTIs) for LAXPC were obtained using \texttt{laxpc\_make\_stdgti}. The 3.0--80.0~keV lightcurves were then extracted using \texttt{laxpc\_make\_lightcurve}. O1, O4, O6, O7 show presence of \tnbs (see Table \ref{tab:burst_fit_params} for reference). In order to study the persistent emission, the GTIs were modified by removing the times when \tnbs were detected. 
The modified GTIs were used again to generate the burst-free lightcurves and spectra, which were later background subtracted. For SXT, a circular region of 15 arcmin radius was chosen and 0.4--6.0~keV lightcurves were extracted using standard \texttt{XSELECT} routines. For the observations where \tnbs were detected, the time segments were removed using \texttt{FILTER TIME} command in \texttt{XSELECT}. For LAXPC data, the 3.0--80.0~keV lightcurves were extracted using \texttt{laxpc\_make\_lightcurve} and the background was subtracted using the standard \texttt{laxpc\_make\_background} tool. 
\begin{figure}[t]
    \centering
    \includegraphics[width=\columnwidth,height=6cm]{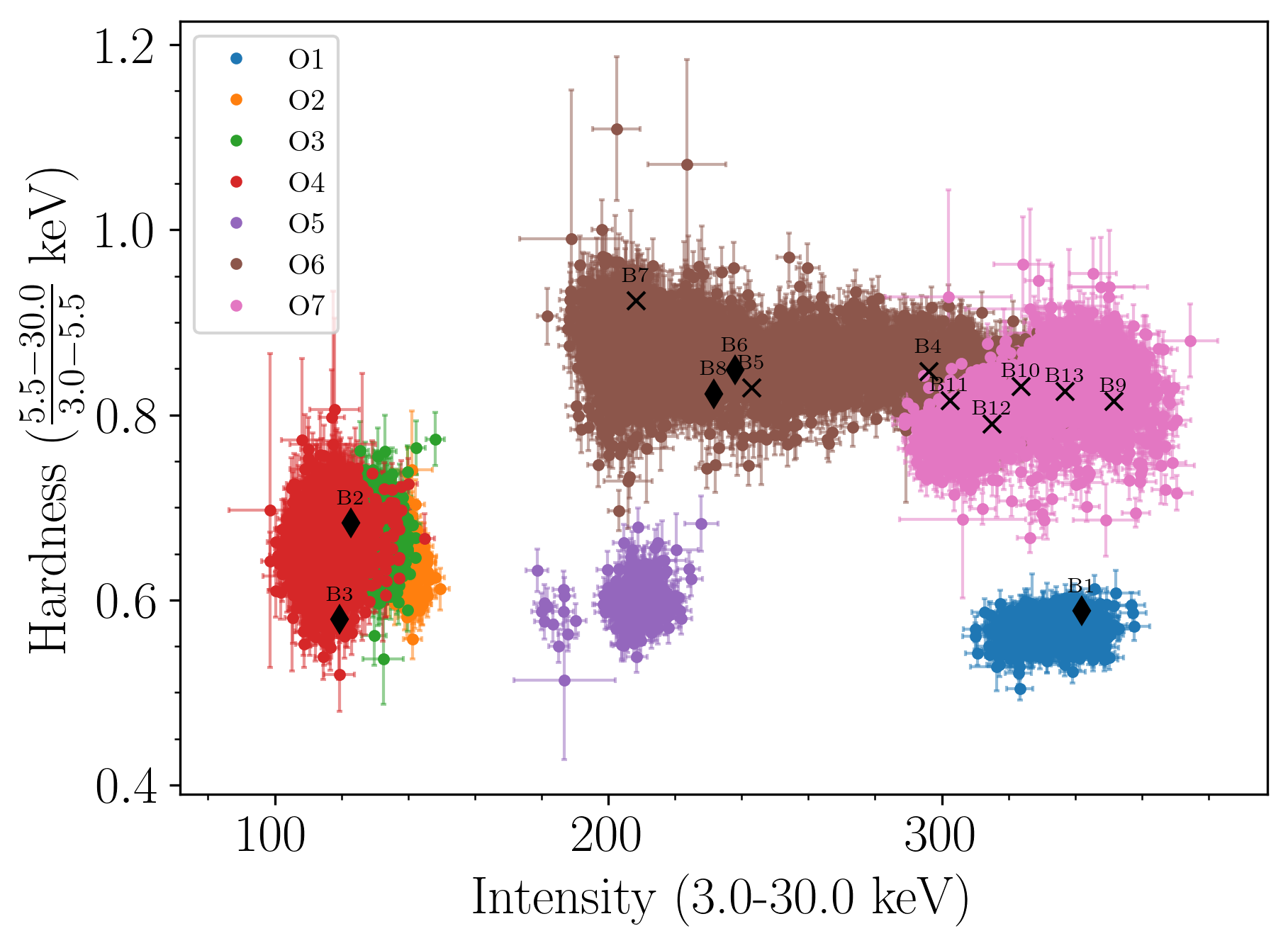}
    \caption{Hardness Intensity Diagram of \source using LAXPC data. The markings are the positions in HID just before the bursts occurred. `$\vardiamondsuit$' represents PRE bursts and `x'  represents Non-PRE bursts.}
    \label{fig:HID}
\end{figure}

To track the spectral evolution, the hardness ratio was calculated from the LAXPC data, as the ratio of the count rates in the 3.0--5.5~keV and 5.5--30.0~keV bands. The HID was constructed using the hardness ratio and the 3--30~keV count rate. The HID was then divided into 25~s bins and the count rates were averaged in each bin. 

From the HID, the hardness ratio varies between $\sim$0.5 to $\sim$1.0 and intensity (count rates) is varying between $\sim$ 100--350~counts~s$^{-1}$. We see \tnbs from 3 distinct regions of the HID. Region 1: where hardness is $\sim$0.6 with high count rates $\geq$300~counts~s$^{-1}$. This region has one burst observed (B1). Region 2: where the hardness ratio is $\sim$0.6 with low count rates $\leq$120~counts~s$^{-1}$ where 2 bursts have been detected (B2 and B3). Region 3: where the hardness ratio is $\geq$0.8 and count rates $\geq$180~counts~s$^{-1}$ where 10 bursts have been detected (B4--B13). 

\begin{table*}[!htbp]
    \centering
    \renewcommand{\arraystretch}{1.5}
    \scriptsize
    \caption{Best-fit parameters for the persistent time-averaged spectrum of all observations.}
    \label{tab:persistent_spec_params}
    \begin{adjustbox}{width=\textwidth,center}
    \begin{tabular}{|l|l|c|c|c|c|c|c|c|}
        \hline
        \textbf{Model} & \textbf{Parameter} & \textbf{O1} & \textbf{O2} & \textbf{O3} & \textbf{O4} & \textbf{O5} & \textbf{O6$^a$} & \textbf{O7} \\
        & & 0.6--25.0~keV & 0.9--25~keV & 0.9--25~keV & 4.0--25.0~keV & 0.9--25~keV & 0.6--25~keV & 0.9--25.0~keV \\  
        \hline
        \texttt{tbabs} 
        & $N_H$ (10$^{22}$ cm$^{-2}$) 
        & $2.27_{-0.15}^{+0.16}$ & $2.24_{-0.12}^{+0.16}$ & $1.67_{-0.17}^{+0.20}$ & $2.50^*$ 
        & $2.41_{-0.19}^{+0.14}$ & $1.91_{-0.06}^{+0.07}$ & $2.12_{-0.12}^{+0.13}$ \\
        \multirow{2}{*}{\texttt{bbody}} 
        & $kT$ (keV) 
        & $1.87 \pm 0.05$ & $1.32_{-0.06}^{+0.05}$ & $1.32 \pm 0.05$ & $1.47_{-0.07}^{+0.08}$ 
        & $1.32_{-0.05}^{+0.07}$ & $1.46 \pm 0.02$ & $1.54 \pm 0.05$ \\
        & norm (10$^{-2}$) 
        & $1.36 \pm 0.06$ & $0.30_{-0.05}^{+0.04}$ & $0.36_{-0.04}^{+0.05}$ & $0.27 \pm 0.06$ 
        & $0.44_{-0.07}^{+0.09}$ & $1.31_{-0.04}^{+0.05}$ & $1.50_{-0.07}^{+0.08}$ \\
        \multirow{2}{*}{\texttt{powerlaw}} 
        & $\Gamma$ 
        & $2.49 \pm 0.03$ & $1.98_{-0.04}^{+0.05}$ & $1.88 \pm 0.05$ & $1.89 \pm 0.07$ 
        & $1.90_{-0.06}^{+0.05}$ & $2.24_{-0.04}^{+0.03}$ & $2.35 \pm 0.03$ \\
        & norm 
        & $1.42_{-0.11}^{+0.13}$ & $0.26_{-0.03}^{+0.04}$ & $0.18 \pm 0.03$ & $0.17_{-0.03}^{+0.04}$ 
        & $0.34_{-0.06}^{+0.05}$ & $0.48 \pm 0.04$ & $1.01 \pm 0.07$ \\
        $F_{p,bol}^1$ 
        & ($\times 10^{-9}$ erg cm$^{-2}$ s$^{-1}$) 
        & $2.64 \pm 0.02$ & $1.12 \pm 0.01$ & $1.06 \pm 0.01$ & $0.96 \pm 0.01$ 
        & $1.74 \pm 0.01$ & $1.82 \pm 0.01$ & $2.58 \pm 0.02$ \\
        $\dot{m}/\dot{M_\mathrm{Edd}}$ 
        &  
        & 0.136 & 0.058 & 0.055 & 0.049 & 0.090 & 0.094 & 0.133 \\
        $\chi^2$/d.o.f 
        &  
        & 1.43/436 & 1.07/427 & 0.84/406 & 1.20/21 & 1.07/476 & 1.24/501 & 1.20/405 \\
        \hline
    \end{tabular}
    \end{adjustbox}
    \vspace{2mm}
\begin{spacing}{0.9} 
\footnotesize{$^1F_{p,bol}$ is total unabsorbed bolometric flux in 3.0--30.0~keV, $^*N_H$ is fixed to $2.5~\times~10^{22}$~cm$^{-2}$ for O4, \\ 
\vspace{-0.5em}
$^a$ up to energies up to 7.0~keV was used for this observation. For the rest of the observations, energies up to 6~keV have been used.}
\end{spacing}
\end{table*}

\subsection{Spectra}\label{sec:per-spec}
Spectral analysis of all the observations are carried out using the time averaged burst-free spectrum from SXT and LAXPC. For LAXPC, the energy range 4.0--25.0~keV is used for spectral studies due to large systematics below 4.0~keV and high background domination above 25.0~keV. For SXT, we have used  0.6--6.0~keV energy range for the spectral fitting. Due to poor data quality in some observations, the lower energy and upper energy limits for the SXT is chosen to be 0.9~keV, and 5.0~keV respectively. Further, occasionally SXT is known to show instrumental artefacts at $\sim$1.8~keV and $\sim$2.4~keV \citep{SXT_singh_et_al}. For observations where the artefacts are found, the energy range 1.6--2.6~keV is excluded for analysis.

The GTIs from both LAXPC and SXT are logically ANDed using \texttt{ftmgtime} to obtain strictly simultaneous broadband spectra. The background spectrum for LAXPC is generated using \texttt{laxpc\_make\_backspectra}. For SXT, the background spectrum \texttt{SkyBkg\_comb\_EL3p5\_Cl\_Rd16p0\_v01.pha} is provided by AstroSat Science Support Cell (ASSC). We then perform joint spectral fitting for LAXPC and SXT using \texttt{XSPEC version: 12.13.0c} \citep{XSPEC}. A gain correction was applied to the SXT data with a fixed slope of 1 and varying offset, to account for the non-linear change in the detector gain. Further, a constant factor was also included for the SXT data to account for the differences in cross-calibration between the SXT and LAXPC instruments. Additionally, a systematic error of 3$\%$ was added to the data as prescribed by the \asat Payload Operations Center team \citep{Bharracharya2017JApA}.

The Tuebingen-Boulder ISM absorption model \texttt{tbabs} was used to account for the absorption by the interstellar medium, which utilizes abundances and updated photoionization cross-sections \citep{Wilms_2000}. 

\begin{figure}
    \centering
    \includegraphics[width=\columnwidth, height=6cm]{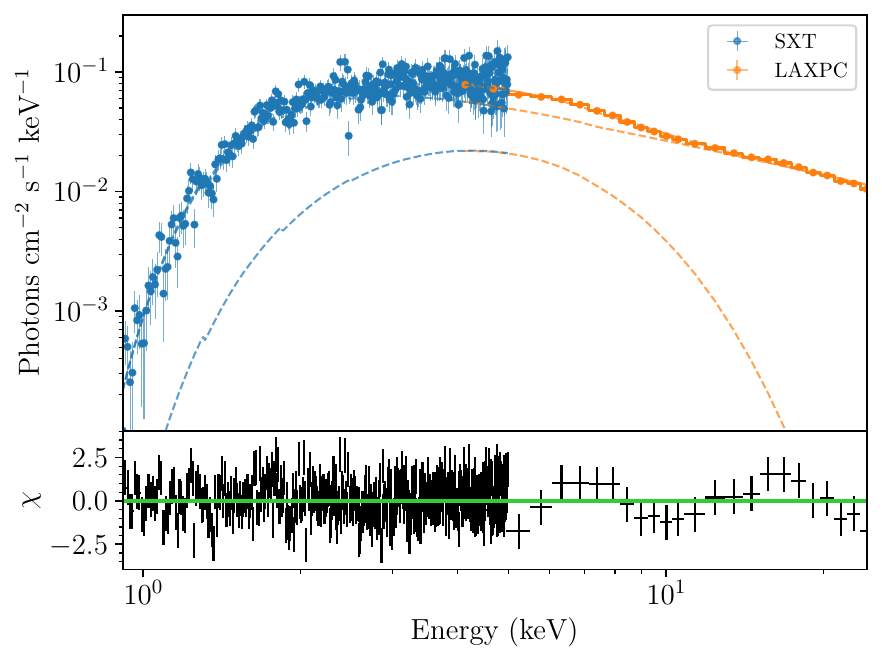}
    \caption{Joint spectral fit of the persistent emission of O2 using SXT (blue) and LAXPC (orange) data with the model combination \texttt{constant*tbabs*(bbody+powerlaw)}}. 
    \label{fig:1904_pers}
\end{figure}

\begin{figure*}
    \centering
    \begin{subfigure}[b]{0.3\textwidth}
        \centering
        \includegraphics[width=\textwidth]{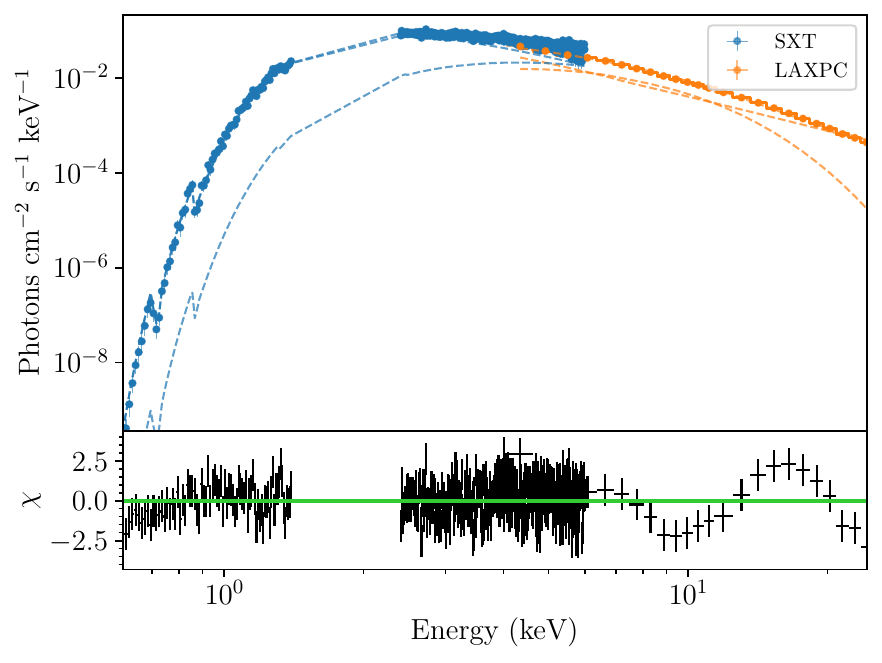}
        \caption{Same as Figure \ref{fig:1904_pers}, for O1}
        \label{fig:O1_fit}
    \end{subfigure}
    \hfill
    \begin{subfigure}[b]{0.3\textwidth}
        \centering
        \includegraphics[width=\textwidth]{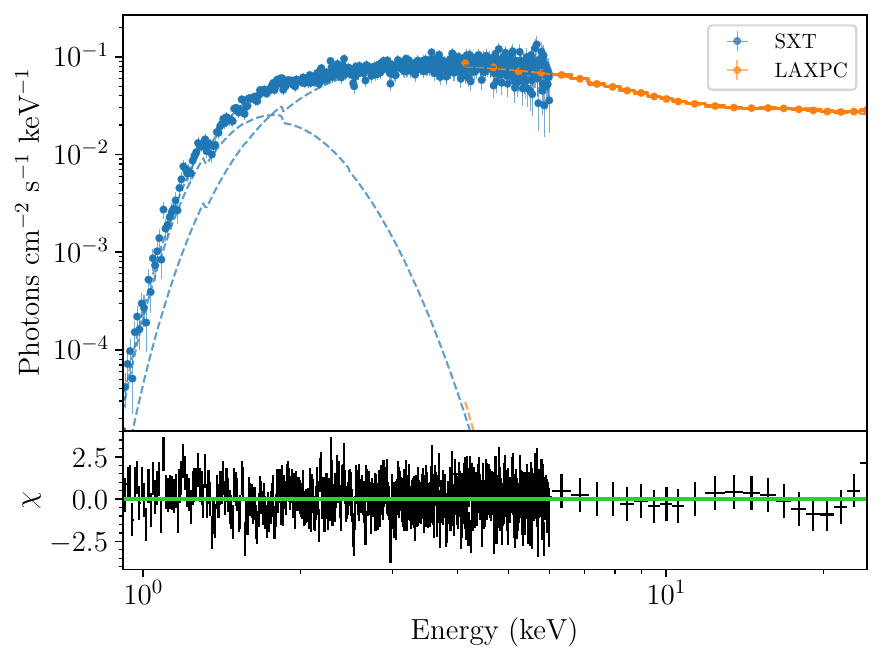}
        \caption{Same as Figure \ref{fig:1904_pers}, for O3}
        \label{fig:O3_fit}
    \end{subfigure}
    \hfill
    \begin{subfigure}[b]{0.3\textwidth}
        \centering
        \includegraphics[width=\textwidth]{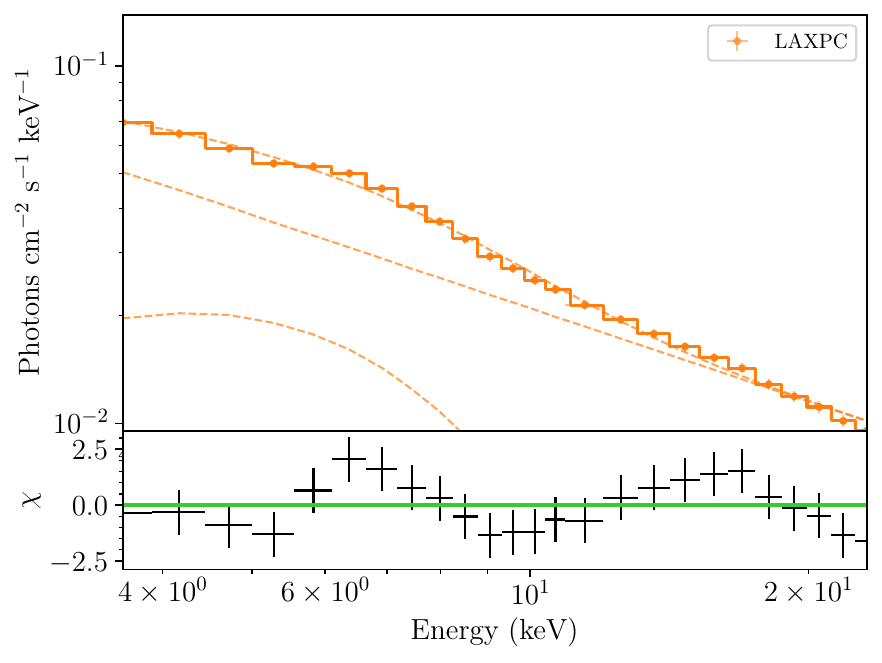}
        \caption{Same as Figure \ref{fig:1904_pers}, for O4}
        \label{fig:O4_fit}
    \end{subfigure}
    \hfill
    \begin{subfigure}[b]{0.3\textwidth}
        \centering
        \includegraphics[width=\textwidth]{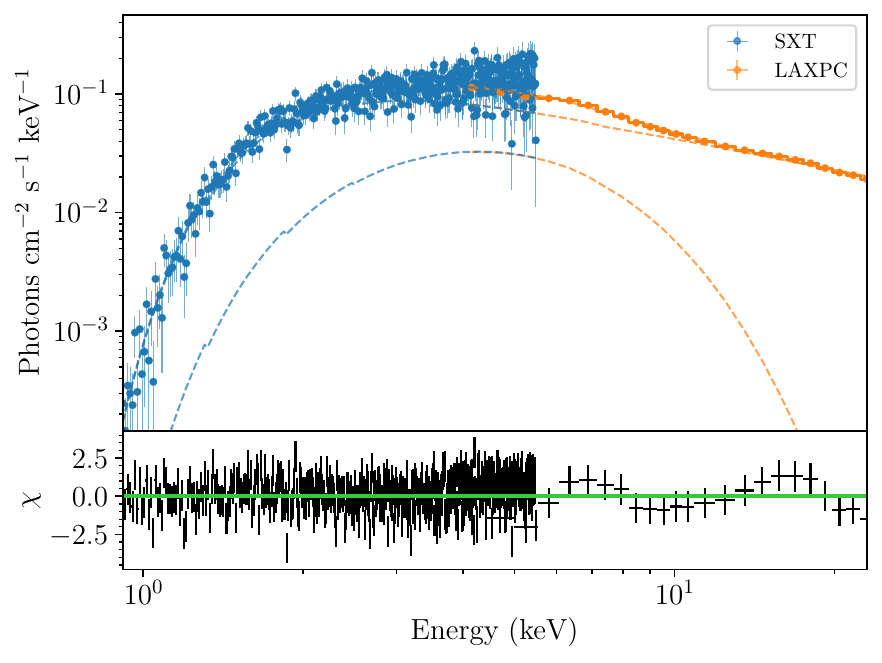}
        \caption{Same as Figure \ref{fig:1904_pers}, for O5}
        \label{fig:O5_fit}
    \end{subfigure}

    \vspace{0.5cm}

    \begin{subfigure}[b]{0.3\textwidth}
        \centering
        \includegraphics[width=\textwidth]{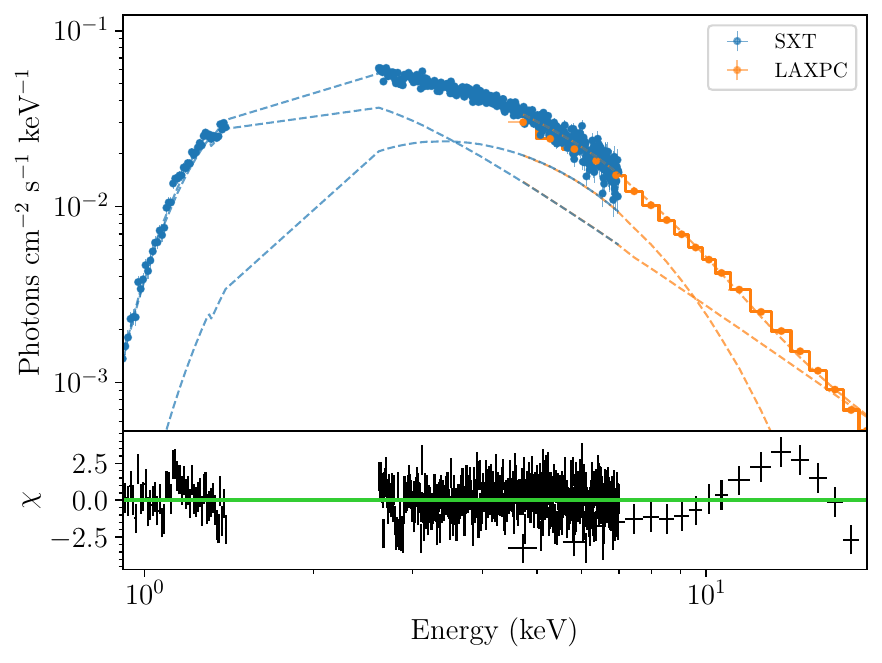}
        \caption{Same as Figure \ref{fig:1904_pers}, for O6}
        \label{fig:O6_fit}
    \end{subfigure}
    \hfill
    \begin{subfigure}[b]{0.3\textwidth}
        \centering
        \includegraphics[width=\textwidth]{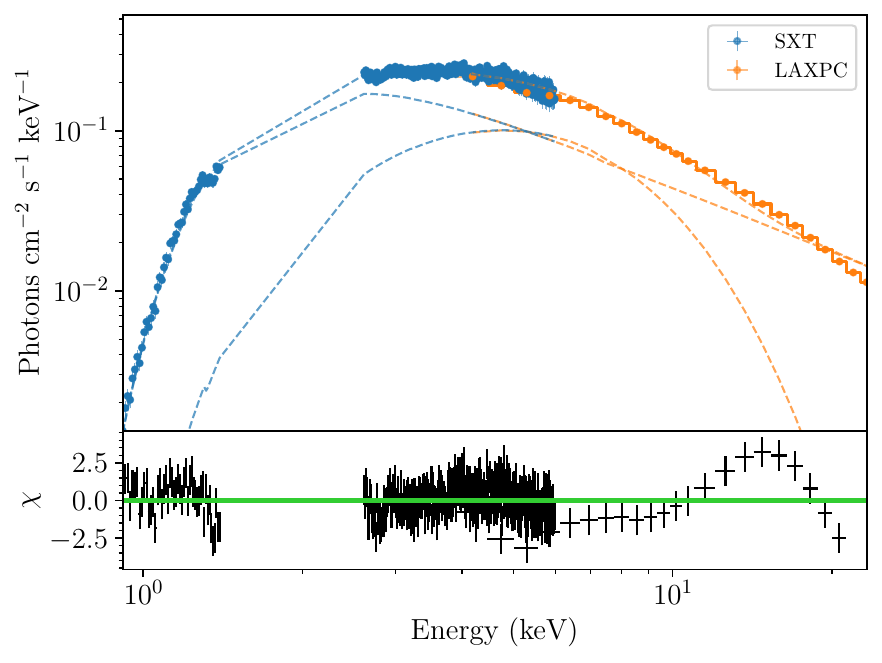}
        \caption{Same as Figure \ref{fig:1904_pers}, for O7}
        \label{fig:O7_fit}
    \end{subfigure}
    \hfill
    \caption{Joint spectral fit}
    \label{fig:all_persistent_fits}
\end{figure*}

The best fit model for the persistent spectrum for all the observations is found to be \texttt{constant*tbabs*(bbody+powerlaw)}. The unabsorbed bolometric flux in the 3.0--30.0~keV energy range is calculated using the \texttt{flux} command. The best fit parameters along with 90\% confidence errors for each of the observations are presented in Table \ref{tab:persistent_spec_params}. The Joint spectral fit along with the residuals are plotted for O2 in Figure \ref{fig:1904_pers}, and for the rest of the observations are in Figure \ref{fig:all_persistent_fits}.

The blackbody temperature ($kT$) is found to be $\sim$1.3--1.9~keV and the powerlaw photon index ($\Gamma$) is $\sim$1.9--2.5. The unabsorbed bolometric flux is found to vary between 0.96--2.64~$\times$10$^{-9}$~erg~cm$^{-2}$~s$^{-1}$, which corresponds to $\sim$5--13\% of the Eddington mass accretion rate. O1 and O7 show higher fluxes, with accretion rates of $\sim$13\% of the Eddington rate. We do not see any disk reflection features around 6~keV in any of the observations.



\subsection{Search for Quasi Periodic Oscillations}
We carried out timing analysis using LAXPC data to search for quasi-periodic variability in the persistent lightcurve. We generated Root Mean Squared (RMS) normalized time averaged power density spectrum (PDS) from 10--2000~Hz using \texttt{laxpc\_find\_freqlag}. The PDS were geometrically binned by a factor of 1.02 in the frequency space using \texttt{laxpc\_rebin\_power}. The LAXPC software normalizes the fractional RMS power (RMS/mean)$^2$/Hz and subtracts out the poisson noise contribution, assuming a deadtime of 42~$\mu$s. This method is repeated for all the observations to investigate rapid X-ray variability. The time averaged power density spectrum of O1 shows the presence of twin kilohertz Quasi Periodic Oscillations (kHz QPOs) at ${964\pm6}$~Hz and ${618\pm10}$~Hz frequencies.

\begin{table}[htbp]
    \centering
    \renewcommand{\arraystretch}{1.2}
    \caption{Best-fit parameters for the time-averaged power density spectrum from 10--2000~Hz reported along with 90\% confidence errors.}
    \resizebox{0.5\textwidth}{!}{ 
    \begin{tabular} {|c|c|c|}
    \hline
    \textbf{Model Component} & \textbf{Parameter} & \textbf{Value}\\
    \hline
    1. \texttt{Lorentzian} & $\sigma$ (Hz) & ${33.27^{+4.04}_{-3.52}}$ \\
                           & $\nu_1$  (Hz) & ${20.28^{+0.98}_{-1.16}}$ \\
                           & Norm (10$^{-3}$) & $9.08\pm0.01$ \\
    2. \texttt{Lorentzian} & $\sigma$ (Hz) & ${125.93^{+32.11}_{-25.35}}$ \\
                           & $\nu_2$  (Hz) &${110.18^{+8.61}_{-10.27}}$ \\
                           & Norm (10$^{-3}$) & $6.19\pm0.01$ \\
    3. \texttt{Lorentzian} & $\sigma$ (Hz) & ${81.84^{+38.71}_{-26.69}}$ \\
        QPO 1 ($\nu_l$)    & $\nu_{3}$  (Hz) & ${618.65^{+9.80}_{-9.60}}$ \\
                           & Norm (10$^{-3}$) & $2.22\pm0.01$\\
    4. \texttt{Lorentzian} & $\sigma$ (Hz) & ${123.62^{+19.01}_{-16.32}}$  \\
        QPO 2 ($\nu_u$)    & $\nu_{3}$  (Hz) & ${964.54^{+5.84}_{-5.86}}$ \\
                           & Norm (10$^{-3}$) & $7.28\pm0.01$ \\
    \hline
                           & $\Delta$\chisq/dof & 254.50/242.0 \\
    \hline
    \end{tabular}
    }
    \label{tab:qpo_params}
\end{table}

These frequencies are the centroid frequencies that were obtained by fitting a zero centred Lorentzian to the power density spectrum, which is given by:

\begin{gather}
 L_\nu = \frac{r^2 \Delta \nu}{2 \pi [(\frac{\Delta \nu}{2})^2 + (\nu - \nu_c)^2]}
\label{eqn:lorentzian}
\end{gather}

Where $\Delta \nu$ is the full width at half maxima (FWHM), $\nu_c$ is the centroid frequency and $r$ is the integrated RMS for the Lorentzian component. The best-fit parameters for the QPOs and 90\% confidence errors obtained are presented in Table \ref{tab:qpo_params}. A total of four Lorentzian were required for adequately fitting the time averaged PDS. In order to quantify the coherence of the QPOs, we calculate the Q-factor ($\nu_c$/$\Delta \nu$), where $\nu_c$ is the centroid frequency and $\Delta\nu$ is the Full Width at Half Maxima, and identify QPOs where Q-factor $>$2. $\nu_l$ and $\nu_u$ together constitute twin kilohertz QPOs. These have been previously reported around similar frequencies by \citep{Mendez2001ApJ, DiSalvo2001ApJ, Wang2018, Anand2024ApJ}. No other observations showed presence of QPOs.

\subsection{Burst Emission and Lightcurves}

\begin{figure}
    \centering
    \includegraphics[width=\columnwidth,height=6cm]{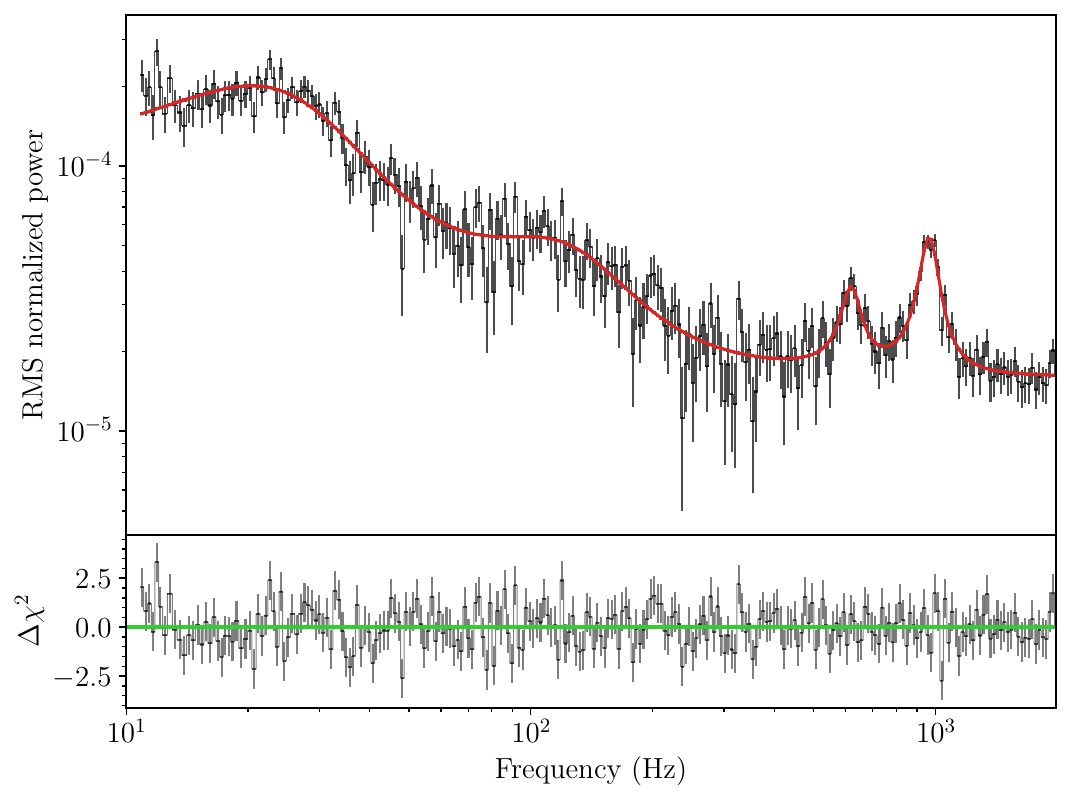}
    \caption{Time averaged power density spectrum of 10--2000 Hz of O1.}
    \label{fig:qpo}
\end{figure}

\begin{figure}
    \centering
    \includegraphics[width=\columnwidth,height=6cm]{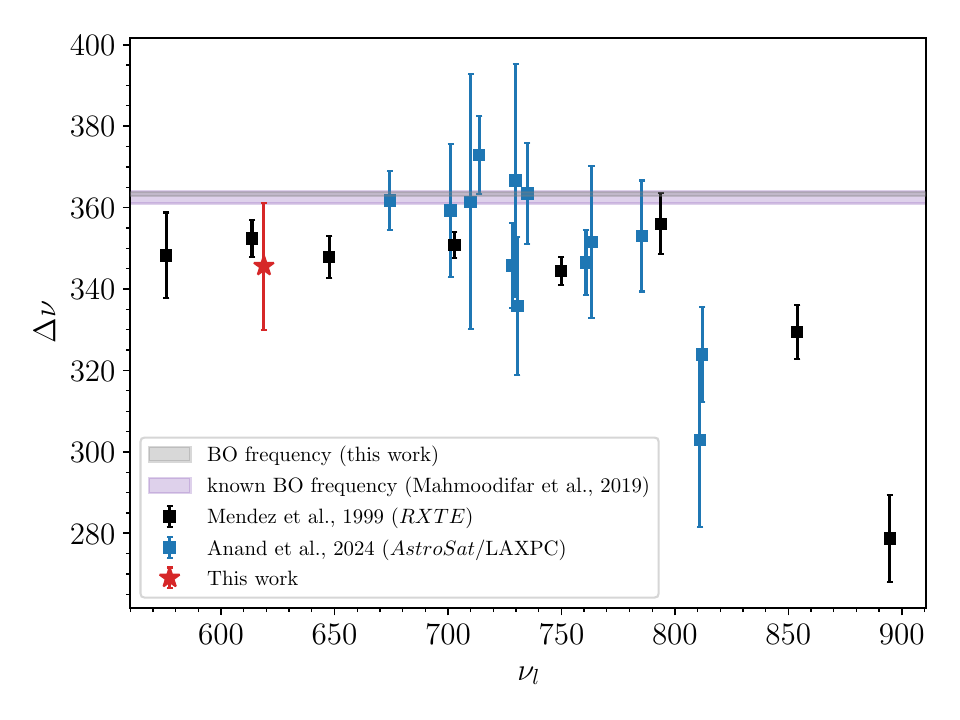}
    \caption{Difference in the upper and lower kHz QPO frequencies $\Delta \nu$ plotted against $\nu_l$. The value of $\Delta \nu$ is smaller than the burst oscillation frequency (see Table \ref{tab:burst_osc})}
    \label{fig:delta_qpo}
\end{figure}

\begin{figure}
    \centering
    \includegraphics[width=\columnwidth,height=6cm]{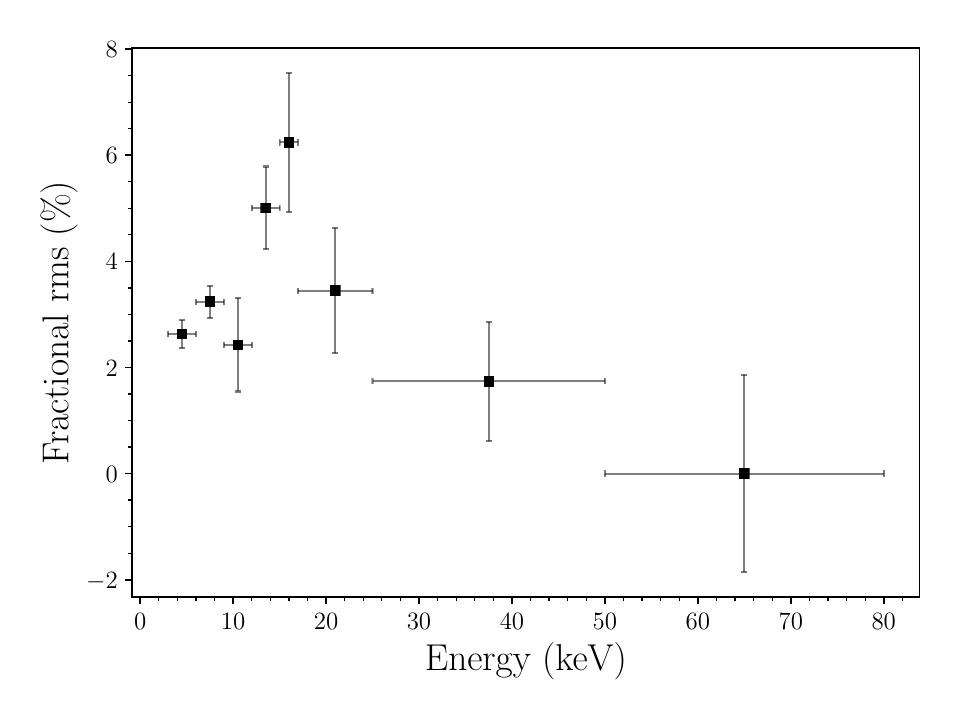}
    \caption{Time averaged RMS spectra of O1 showing the energy dependence of the upper kHz QPO.}
    \label{fig:rms_spectra}
\end{figure}

In order to understand the energy dependence of the upper kHz QPO, we plot RMS as a function of energy (see Figure \ref{fig:rms_spectra}). From the RMS spectra, we see that the upper QPOs are strongest in the 14.0--16.0~keV energy range, with a peak fractional rms of $6.24\pm1.31$\% at $16.0\pm1.0$~keV. We also see an increase in amplitude till $\sim$16~keV, beyond which it starts decreasing. This may indicate that $\sim$ 16~keV photons contributed relatively more in QPOs than other energy band photons.

\begin{table*}
    \centering
    \renewcommand{\arraystretch}{1.2}
    \scriptsize
    \caption{Best-fit parameters of LAXPC 3.0--80.0~keV burst light curve fitting. }
    \begin{adjustbox}{width=0.7\textwidth, center}
    \begin{tabular} {|c|c|c|c|c|c|c|c|}
    \hline
    \textbf{Burst} & \textbf{Peak Count} & \textbf{Rise time} & \textbf{Decay time} & \textbf{t$_{90}^*$} & \textbf{PRE} \\
    & \textbf{rate~\countsps} & \textbf{(s)} & \textbf{(s)} & \textbf{(s)} & \\
    \hline
    B1 & 5962 $\pm$ 77 & 0.53 $\pm$ 0.11 & 6.08 $\pm$ 0.38 & 11.75 $\pm$ 0.39 & y \\
    B2 & 3838 $\pm$ 61 & 0.38 $\pm$ 0.14 & 4.59 $\pm$ 0.41 & 12.15 $\pm$ 0.43 & y  \\
    B3 & 3756 $\pm$ 61 & 0.90 $\pm$ 0.17 & 6.02 $\pm$ 0.60 & 12.38 $\pm$ 0.62 & y \\
    B4 & 4398 $\pm$ 66 & 0.22 $\pm$ 0.03 & 6.58 $\pm$ 0.18 & 12.33 $\pm$ 0.18 & n \\
    B5 & 4726 $\pm$ 68 & 0.15 $\pm$ 0.12 & 6.93 $\pm$ 0.35 & 12.28 $\pm$ 0.37 & n \\
    B6 & 5594 $\pm$ 74 & 0.30 $\pm$ 0.04 & 5.57 $\pm$ 0.17 & 11.94 $\pm$ 0.17 & y \\
    B7 & 5404 $\pm$ 73 & 0.51 $\pm$ 0.10 & 5.04 $\pm$ 0.28 & 12.47 $\pm$ 0.30 & n \\
    B8 & 4684 $\pm$ 68 & 0.47 $\pm$ 0.05 & 4.88 $\pm$ 0.24 & 12.01 $\pm$ 0.24 & y\\
    B9 & 3932 $\pm$ 62 & 0.37 $\pm$ 0.08 & 6.19 $\pm$ 0.45 & 13.06 $\pm$ 0.46 & n \\
    B10 & 5570 $\pm$ 74 & 0.57 $\pm$ 0.08 & 4.80 $\pm$ 0.34 & 12.11 $\pm$ 0.35 & n \\
    B11 & 5588 $\pm$ 74 & 0.61 $\pm$ 0.10 & 4.36 $\pm$ 0.39 & 12.37 $\pm$ 0.40 & n \\
    B12 & 6230 $\pm$ 78 & 0.31 $\pm$ 0.11 & 5.99 $\pm$ 0.42 & 11.63 $\pm$ 0.43 & n \\
    B13 & 5822 $\pm$ 76 & 0.54 $\pm$ 0.11 & 5.74 $\pm$ 0.32 & 11.85 $\pm$ 0.34 & n \\
    \hline
    \end{tabular}
    \end{adjustbox}
    \footnotesize{$^*$ t$_{90}$ is a measure to quantify the duration during which 90\% of the total burst energy (or counts) is emitted.}
    \label{tab:burst_fit_params}.
\end{table*}
We further characterize the difference between the upper and lower kHz QPO frequencies. Figure \ref{fig:delta_qpo} is a plot representing, $\Delta \nu$ i.e. $\nu_u - \nu_l$ as a function of $\nu_l$. The  $\Delta \nu$ value is found to be $345.64\pm 15.59$~Hz, which is smaller than the frequency of burst  oscillations seen in this source.

The observations O1, O4, O6 and O7 show the presence of a total of 13 \tnbs. The burst lightcurves are modelled using \texttt{mpfit}\footnote{\url{https://github.com/segasai/astrolibpy/blob/master/mpfit/mpfit.py}} module which uses Levenberg-Marquardt least square minimization to fit the burst lightcurves. The best-fit burst parameters from the lightcurve are given in Table \ref{tab:burst_fit_params}. While the observations O1 and O4 are located in region 1 and region 2 of the HID respectively, (see Figure \ref{fig:HID}), a total of 3 \tnbs (B1, B2 and B3) have been reported in these two observations. The peak count rate for B1 is $\sim$6038~\countsps, while it is $\sim$3700~\countsps for bursts B2 and B3. Other parameters such as the burst duration, time taken to reach the peak count rate and decay rate are comparable for these three bursts. The Observations O6 and O7 belonging to region R3 in the HID shows a total of 11 \tnbs, with peak count rates raging from $\sim$3900~\countsps (for B9) to $\sim$6200~\countsps (for B12), which is the highest peak count rate among all the bursts.

Burst B9 and B11, though located in the same region in the HID, show a drastic difference in the time taken to peak, i.e. $\sim$2~s, and $\sim$0.3~s, respectively. This contrast is also seen in burst duration. While B9 lasts for $\sim$17~s, B11 lasts for $\sim$10~s.

\begin{figure}
    \centering
    \includegraphics[width=\columnwidth,height=6cm]{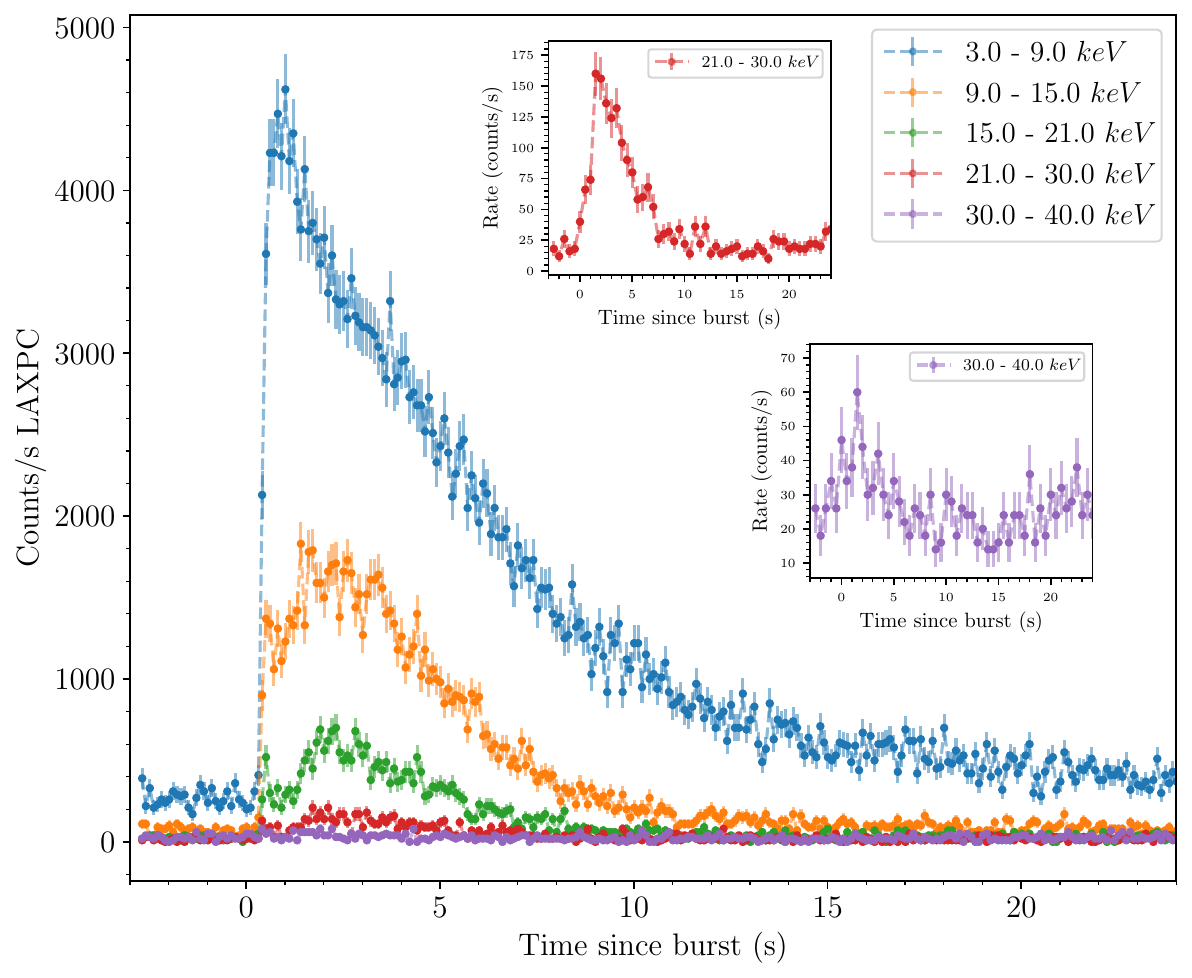}
    \caption{Energy resolved burst lightcurve for burst B12. The lightcurves of 21.0--30.0~keV and 30.0--50.0~keV energy range are shown separately, binned at 0.5~s.}
    \label{fig:energy_resolved_burst}
\end{figure}

\begin{table*}
    \centering
    \scriptsize
    \renewcommand{\arraystretch}{1.2} 
    \caption{Burst spectral parameters}
    \begin{adjustbox}{width=0.7\textwidth, center}
    \begin{tabular} {|c|c|c|c|c|c|c|}
    \hline
    \textbf{Burst} & \textbf{\boldmath$kT_{peak}$} & \textbf{\boldmath$F_{peak, bol}$} & \textbf{\boldmath$R_{TD}$} & \textbf{\boldmath$L/L_\mathrm{Edd}$} & \textbf{\boldmath$\dot{M}/\dot{M}_\mathrm{Edd}$} \\
    \hline

B1 & $2.70\pm0.09$ & $5.94_{-0.16}^{+0.09}$ & $4.70_{-0.29}^{+0.10}$ & 1.09 & 3.06 \\
B2 & $2.91\pm0.11$ & $3.55_{-0.11}^{+0.09}$ & $3.49_{-0.24}^{+0.09}$ & 0.65 & 1.83 \\
B3 & $3.09\pm0.11$ & $3.38_{-0.10}^{+0.08}$ & $3.18_{-0.22}^{+0.08}$ & 0.62 & 1.74 \\
B4 & $2.47\pm0.10$ & $3.37_{-0.11}^{+0.06}$ & -- & 0.62 & 1.73 \\
B5 & $2.61\pm0.09$ & $4.21_{-0.13}^{+0.07}$ & -- & 0.77 & 2.17 \\
B6 & $2.65\pm0.08$ & $5.48_{-0.12}^{+0.10}$ & $5.54_{-0.33}^{+0.10}$ & 1.01 & 2.82 \\
B7 & $2.71_{-0.09}^{+0.10}$ & $5.28_{-0.16}^{+0.09}$ & -- & 0.97 & 2.72 \\
B8 & $2.63\pm0.10$ & $4.12_{-0.12}^{+0.09}$ & $5.12_{-0.35}^{+0.09}$ & 0.76 & 2.12 \\
B9 & $2.37\pm0.10$ & $2.95_{-0.08}^{+0.09}$ & -- & 0.54 & 1.52 \\
B10 & $2.70\pm0.09$ & $5.26_{-0.16}^{+0.07}$ & -- & 0.97 & 2.71 \\
B11 & $2.65\pm0.08$ & $5.17_{-0.15}^{+0.10}$ & -- & 0.95 & 2.66 \\
B12 & $2.69\pm0.08$ & $5.56_{-0.15}^{+0.12}$ & -- & 1.02 & 2.86 \\
B13 & $2.71\pm0.08$ & $5.65_{-0.16}^{+0.10}$ & -- & 1.04 & 2.91 \\

    \hline
    \end{tabular}
    \end{adjustbox}
    \footnotesize{$kT_{peak}$ is the maximum temperature achieved during the burst, $F_{peak, bol}$ is the maximum bolometric flux attained during the burst (in $\times 10^{-9}$~erg~cm$^{-2}$~s$^{-1}$), $R_{TD}$ is the touchdown radius in km, $L/L_\mathrm{Edd}$ is the ratio of Source peak luminosity to Eddington luminosity $L_\mathrm{Edd}$ i.e. $1.76~\times~10^{38}$~erg~s$^{-1}$, $\dot{M}/\dot{M}_\mathrm{Edd}$ is the mass accretion rate at the peak of the burst relative to the Eddington limit $\dot{M}_\mathrm{Edd}$ i.e. $8.8~\times 10^4$~g~cm$^{-2}$~s$^{-1}$. $L_\mathrm{Edd}$  and $\dot{M}_\mathrm{Edd}$ are calculated for a NS of mass $1.4~M_\odot$ and radius of 10~km \citep{Galloway2008ApJS}.}
    \label{tab:burst_spec_params}
\end{table*}
Energy resolved lightcurves of $\sim$0.1~s are extracted in the energy ranges 3.0--9.0~keV (E1), 9.0--15.0~keV (E2), 15.0--21.0~keV (E3), 21.0--30.0~keV (E4) and 30.0--40.0~keV (E5). We observe that all the bursts show the presence of photons $>$20~keV. In addition, the bursts except B4, B6 and B8 show photons of energy $>30$~keV in the energy resolved lightcurves as seen in Figure \ref{fig:energy_resolved_burst}. The energy resolved burst lightcurves of all the bursts are presented in Figure \ref{fig:all_energy_resolved_bursts}. This behaviour is further investigated using spectroscopy in Section \ref{sec:spec}.

\subsection{Time resolved burst spectroscopy}\label{sec:spec}
We extract the LAXPC spectra for each of the bursts. The start time is considered as the time when the count rate exceeds the persistent emission by 10$\%$ of the peak count rate of the burst. To investigate the spectral evolution, consecutive spectra of 0.5~s from the start of the burst were extracted. The SXT instrument was not used for the analysis due to its larger time resolution of 2.37~s. Following several works such as \citep{Mondal2017MNRAS, DiSalvo2000ApJ,Falanga2006AA}, we have used the persistent emission spectrum before or after the corresponding burst as the background spectrum in order to eliminate the contamination of burst-free photons. 

For each interval of the time resolved burst spectra, we used \texttt{XSPEC version: 12.13.0c} \citep{XSPEC} to fit the spectrum. We accounted for the interstellar medium absorption using the model \texttt{tbabs} and froze the N$_{H}$ value to 2.5 $\times$ 10$^{22}$~cm$^{-2}$  \citep{Mondal2017MNRAS} since we are not statistically capable of constraining this parameter using LAXPC. 

\begin{figure}
    \centering \includegraphics[width=\columnwidth,height=8.5cm]{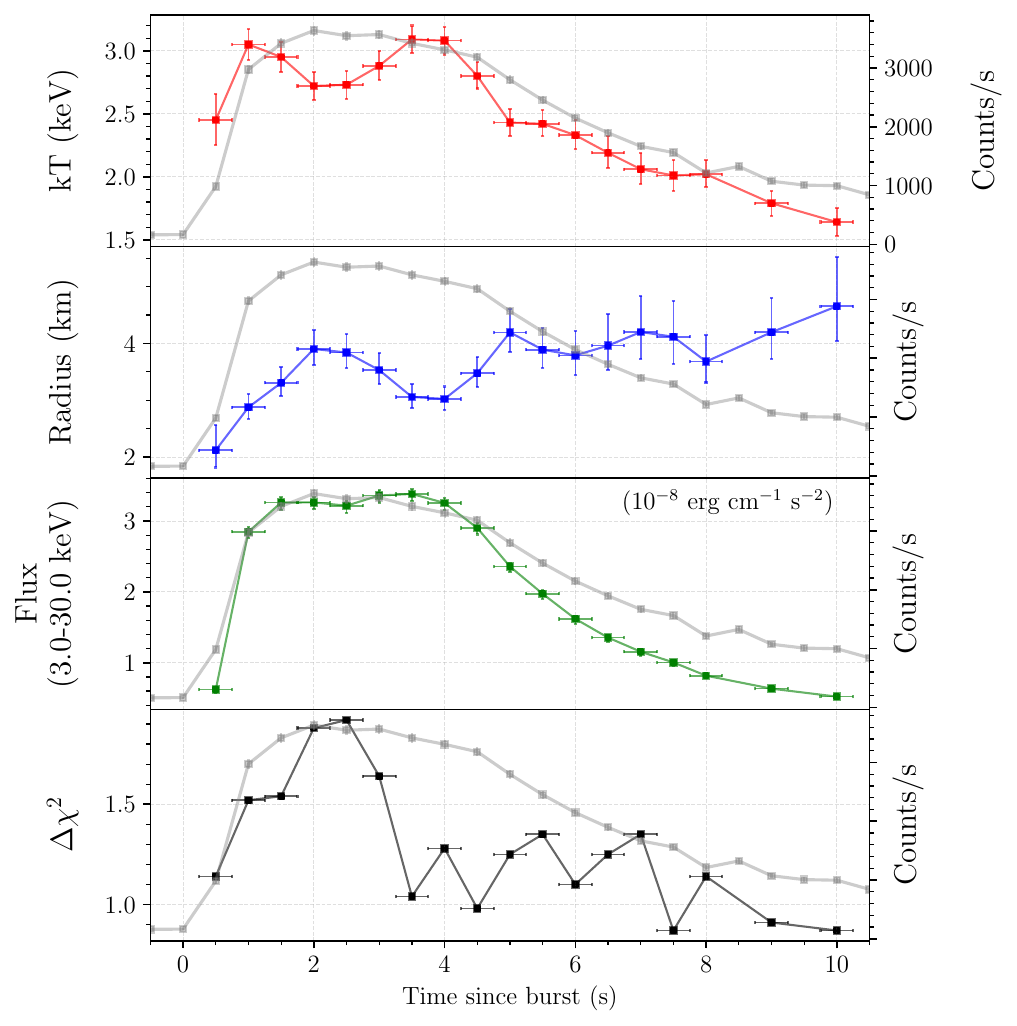}
    \caption{Evolution of spectral parameters of burst B2, elucidating PRE. The top panel shows the evolution of blackbody temperature $kT$~(keV), second panel shows the evolution of blackbody radius R$_{bb}$~(in km), third panel showing burst flux and bottom panel showing fit statistic}
    \label{fig:time_resolved}
\end{figure}

Conventionally, burst spectra are fitted with a thermal blackbody model. We note here, that we do not adopt the variable flux ($f_a$) method, as we discuss later in Section \ref{sec:PREdisc}, and we adopt the conventional method for the rest of this work. Figure \ref{fig:time_resolved} shows the evolution of burst spectral parameters. 

We adopted a similar method as done by \citep{Guver2012ApJ}, for the identification of PRE bursts, estimating touchdown radius $R_{TD}$ and touchdown flux $R_{TD}$. We infer that B1, B2, B3, B6, and B8 are indicative of PRE phenomena.

\begin{figure*}
    \centering
    \begin{subfigure}[b]{0.3\textwidth}
        \centering
        \includegraphics[width=\textwidth]{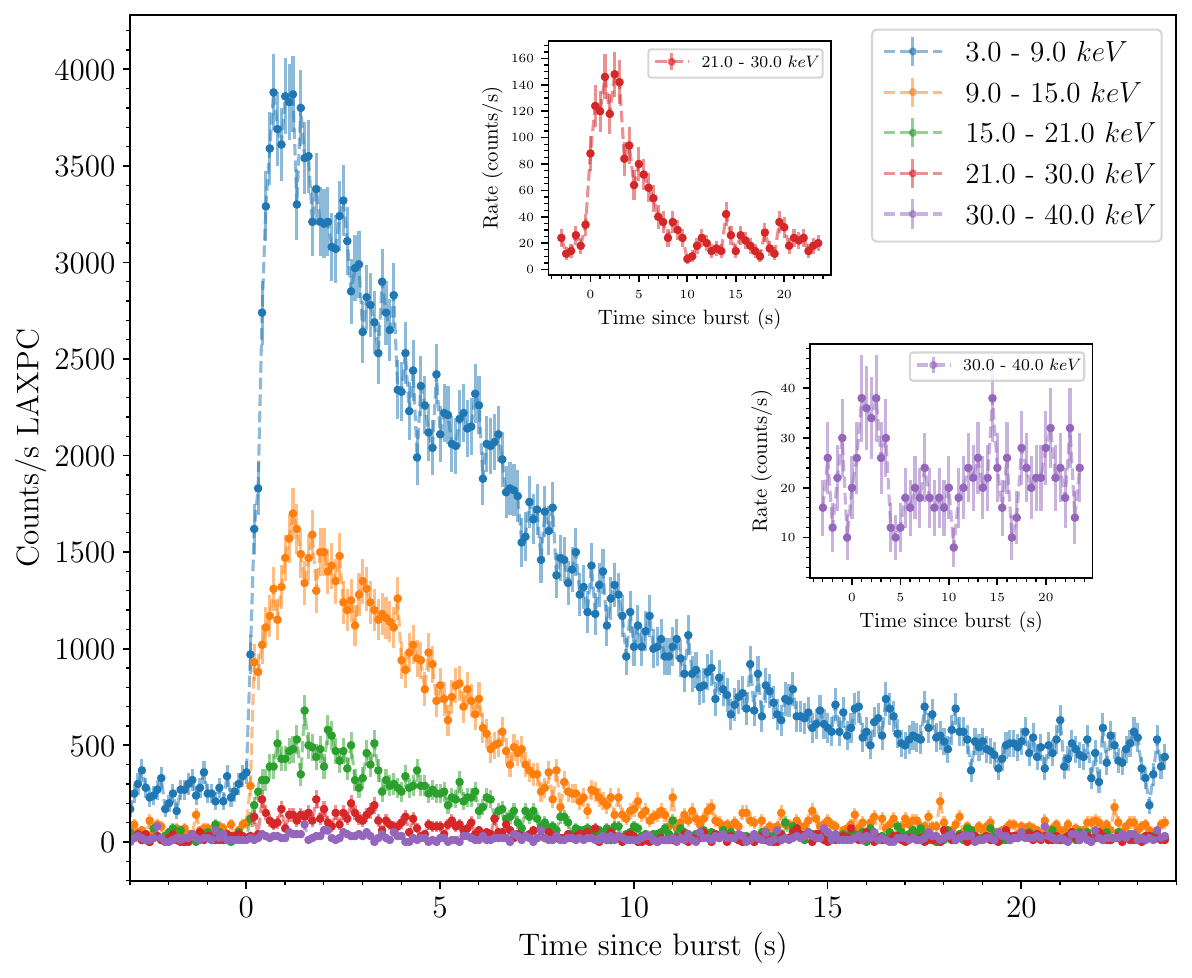}
        \caption{Same as Figure \ref{fig:energy_resolved_burst}, for B1}
        \label{fig:energy_resolved_lc_b1}
    \end{subfigure}
    \hfill
    \begin{subfigure}[b]{0.3\textwidth}
        \centering
        \includegraphics[width=\textwidth]{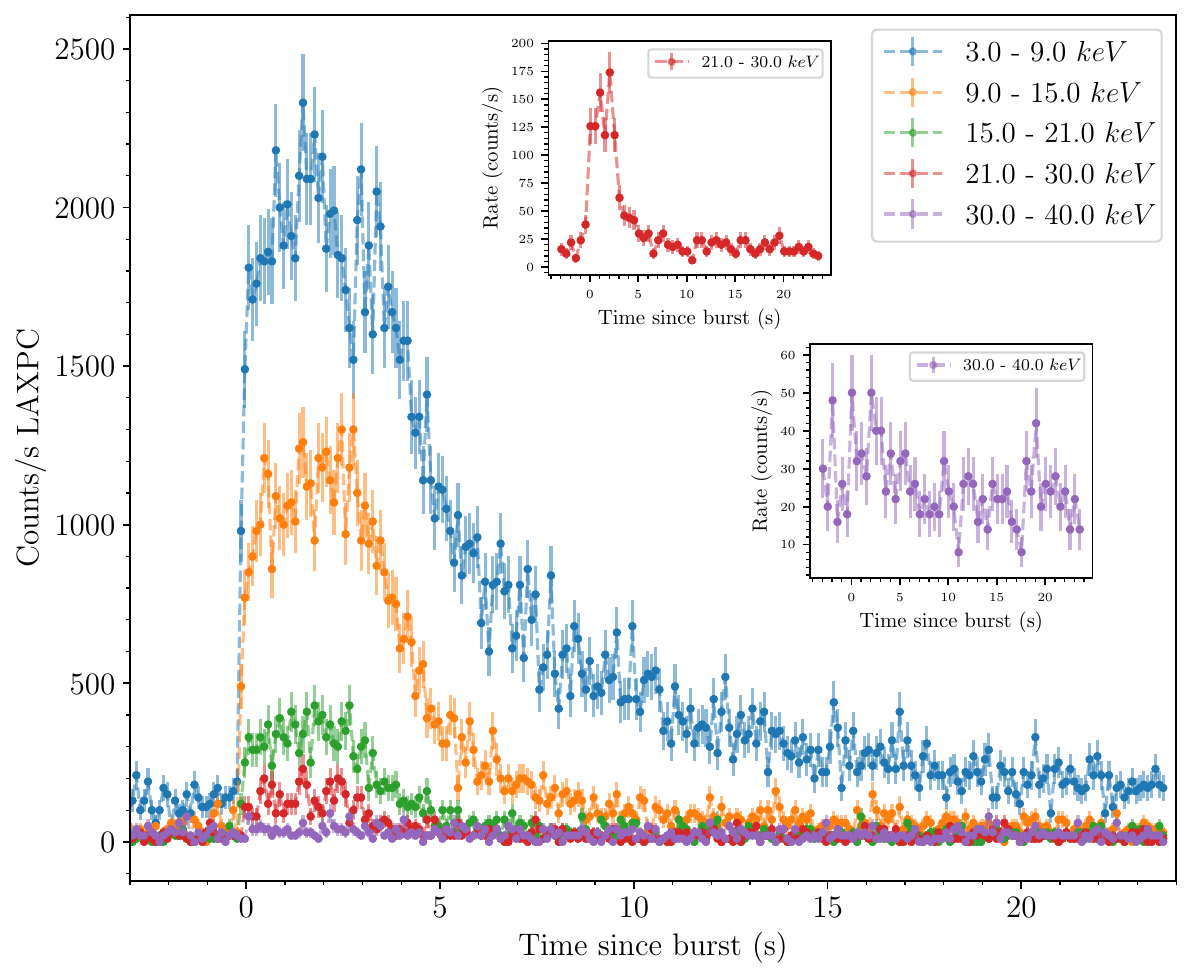}
        \caption{Same as Figure \ref{fig:energy_resolved_burst}, for B2}
        \label{fig:energy_resolved_lc_b2}
    \end{subfigure}
    \hfill
    \begin{subfigure}[b]{0.3\textwidth}
        \centering
        \includegraphics[width=\textwidth]{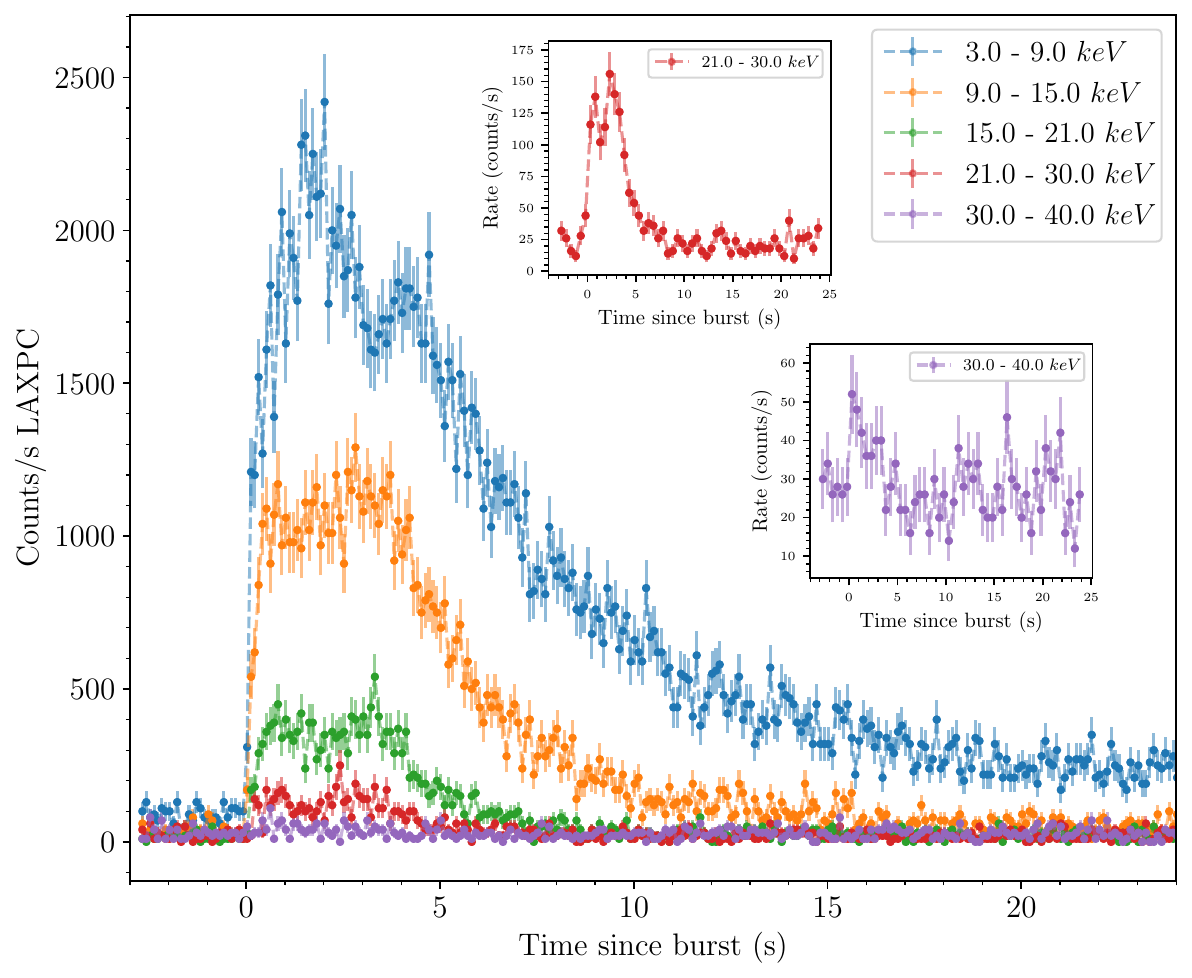}
        \caption{Same as Figure \ref{fig:energy_resolved_burst}, for B3}
        \label{fig:energy_resolved_lc_b3}
    \end{subfigure}

    \vspace{0.5cm}

    \begin{subfigure}[b]{0.3\textwidth}
        \centering
        \includegraphics[width=\textwidth]{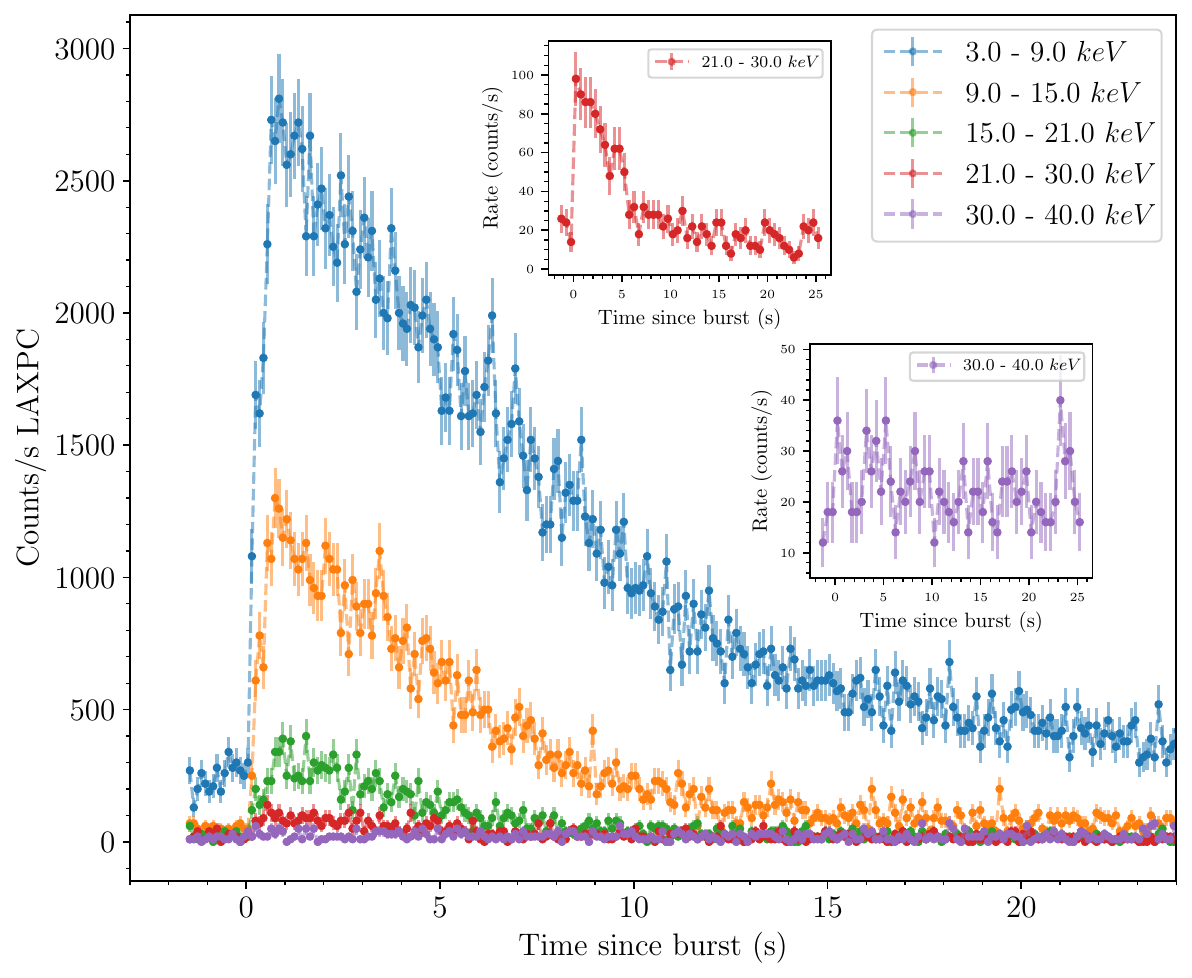}
        \caption{Same as Figure \ref{fig:energy_resolved_burst}, for B4}
        \label{fig:energy_resolved_lc_b4}
    \end{subfigure}
    \hfill
    \begin{subfigure}[b]{0.3\textwidth}
        \centering
        \includegraphics[width=\textwidth]{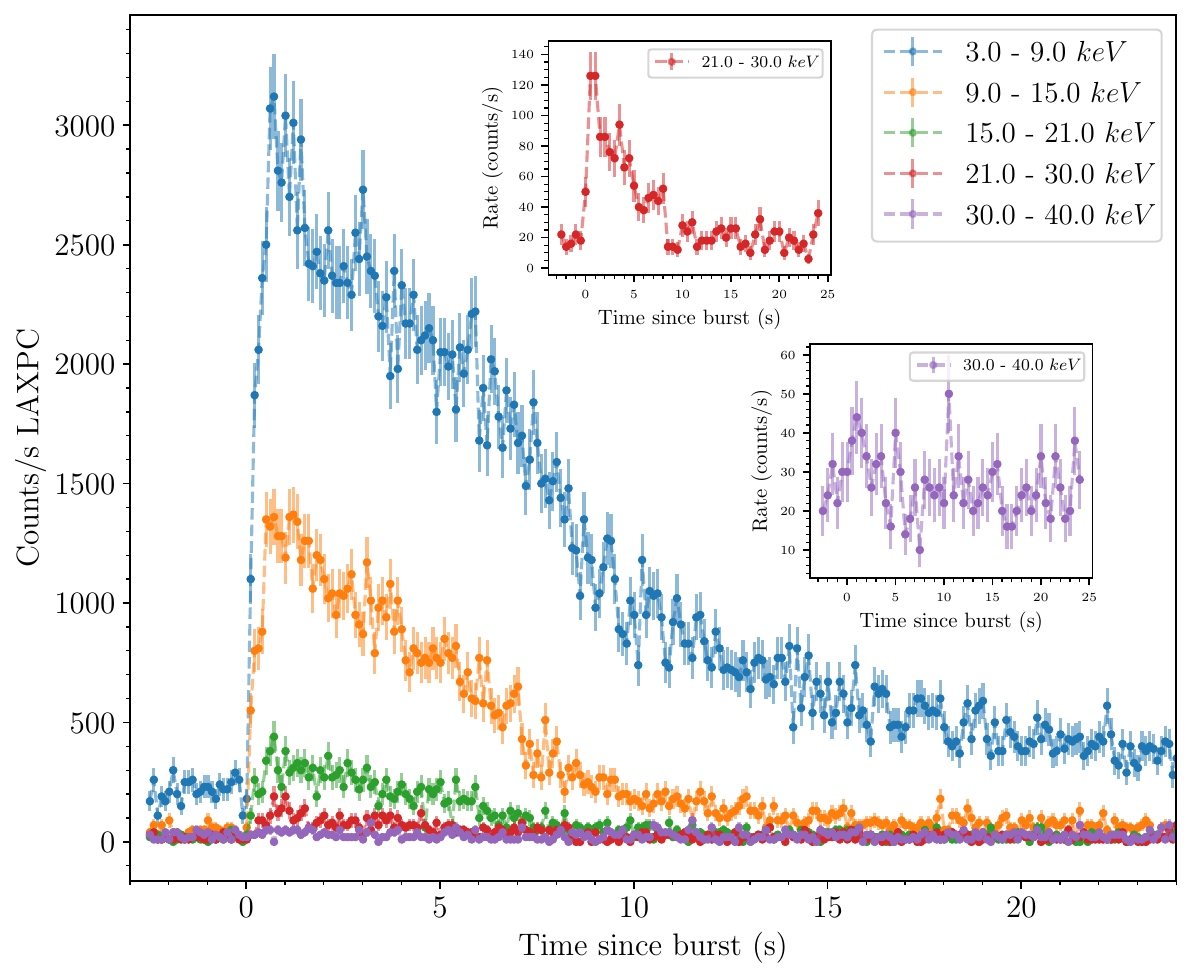}
        \caption{Same as Figure \ref{fig:energy_resolved_burst}, for B5}
        \label{fig:energy_resolved_lc_b5}
    \end{subfigure}
    \hfill
    \begin{subfigure}[b]{0.3\textwidth}
        \centering
        \includegraphics[width=\textwidth]{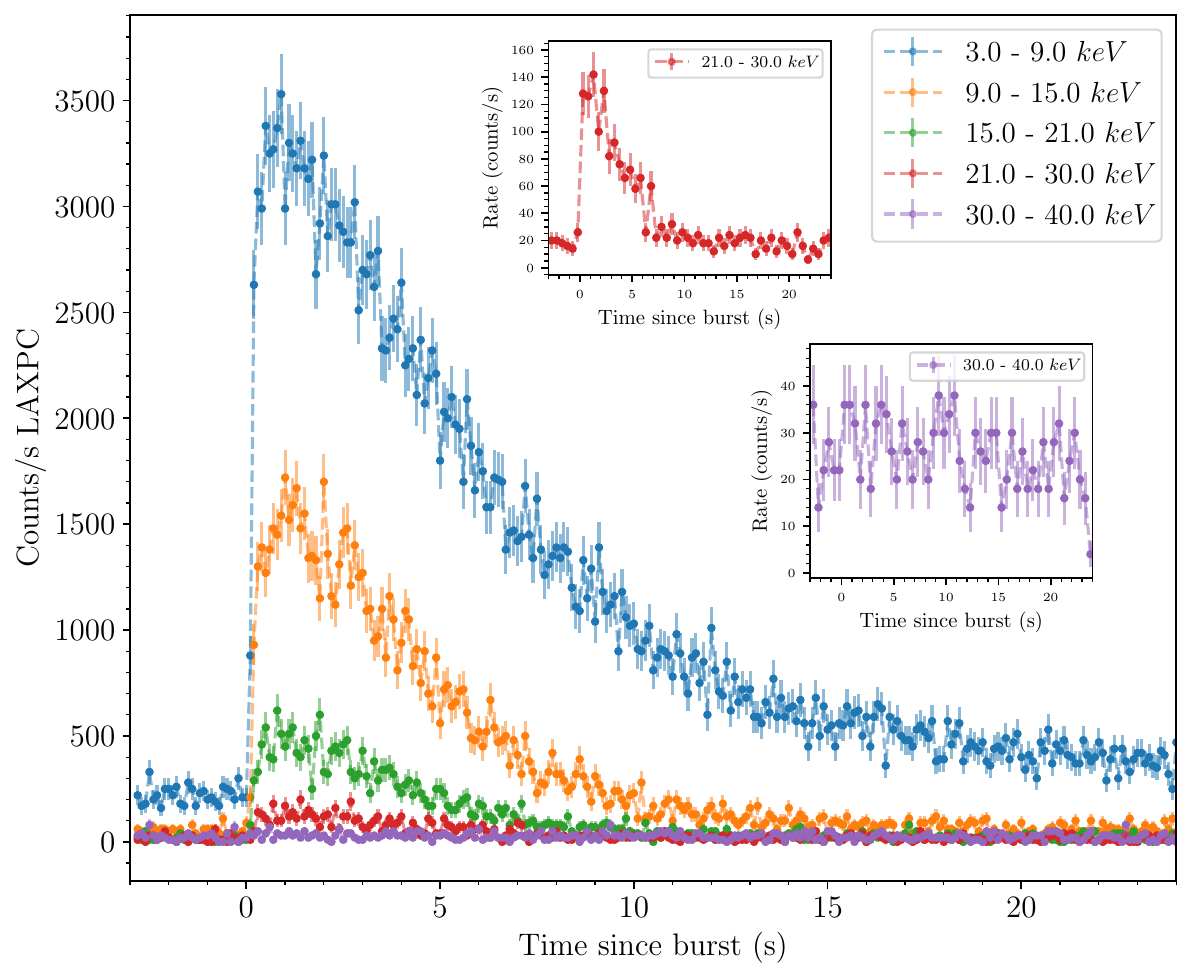}
        \caption{Same as Figure \ref{fig:energy_resolved_burst}, for B6}
        \label{fig:energy_resolved_lc_b6}
    \end{subfigure}

    \vspace{0.5cm}

    \begin{subfigure}[b]{0.3\textwidth}
        \centering
        \includegraphics[width=\textwidth]{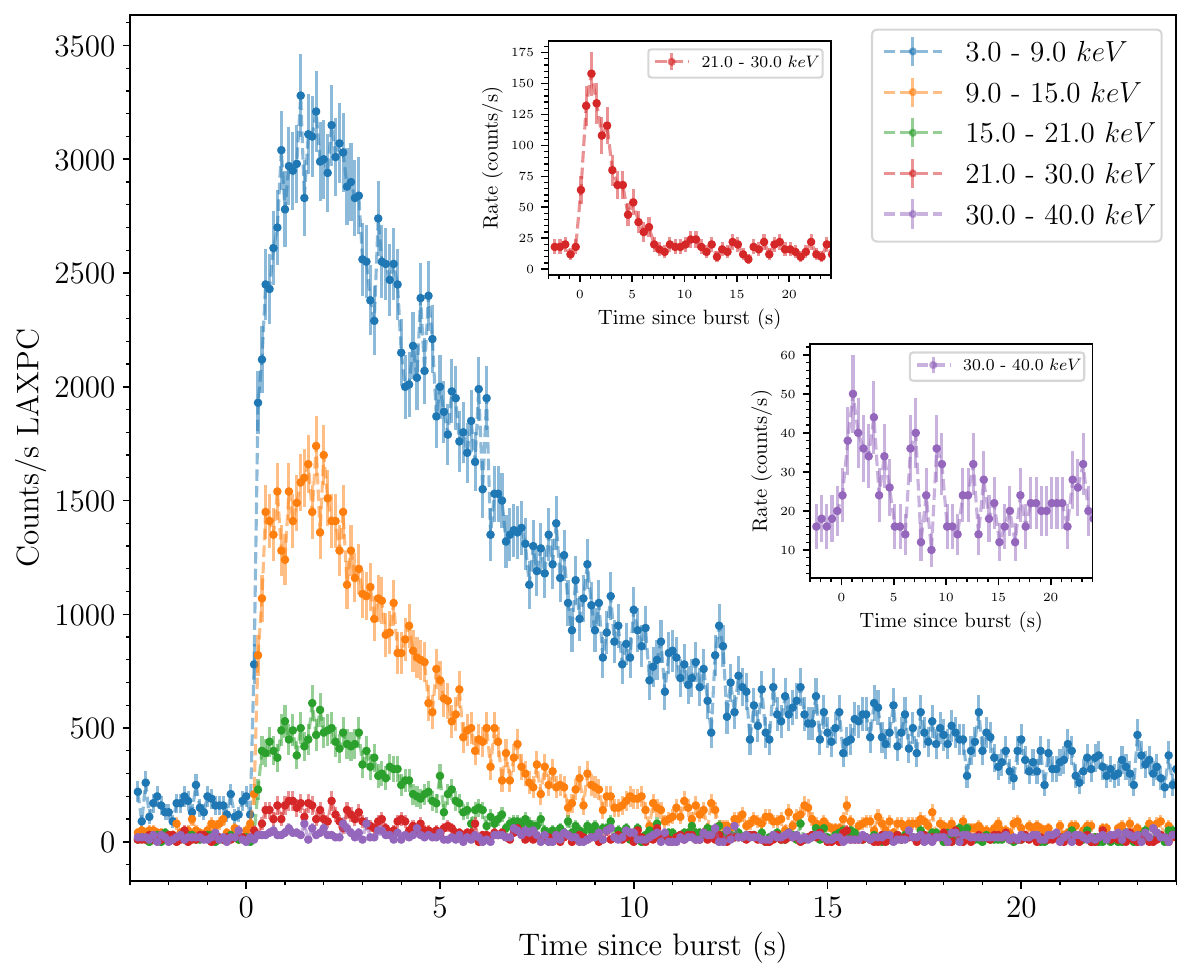}
        \caption{Same as Figure \ref{fig:energy_resolved_burst}, for B7}
        \label{fig:energy_resolved_lc_b7}
    \end{subfigure}
    \hfill
    \begin{subfigure}[b]{0.3\textwidth}
        \centering
        \includegraphics[width=\textwidth]{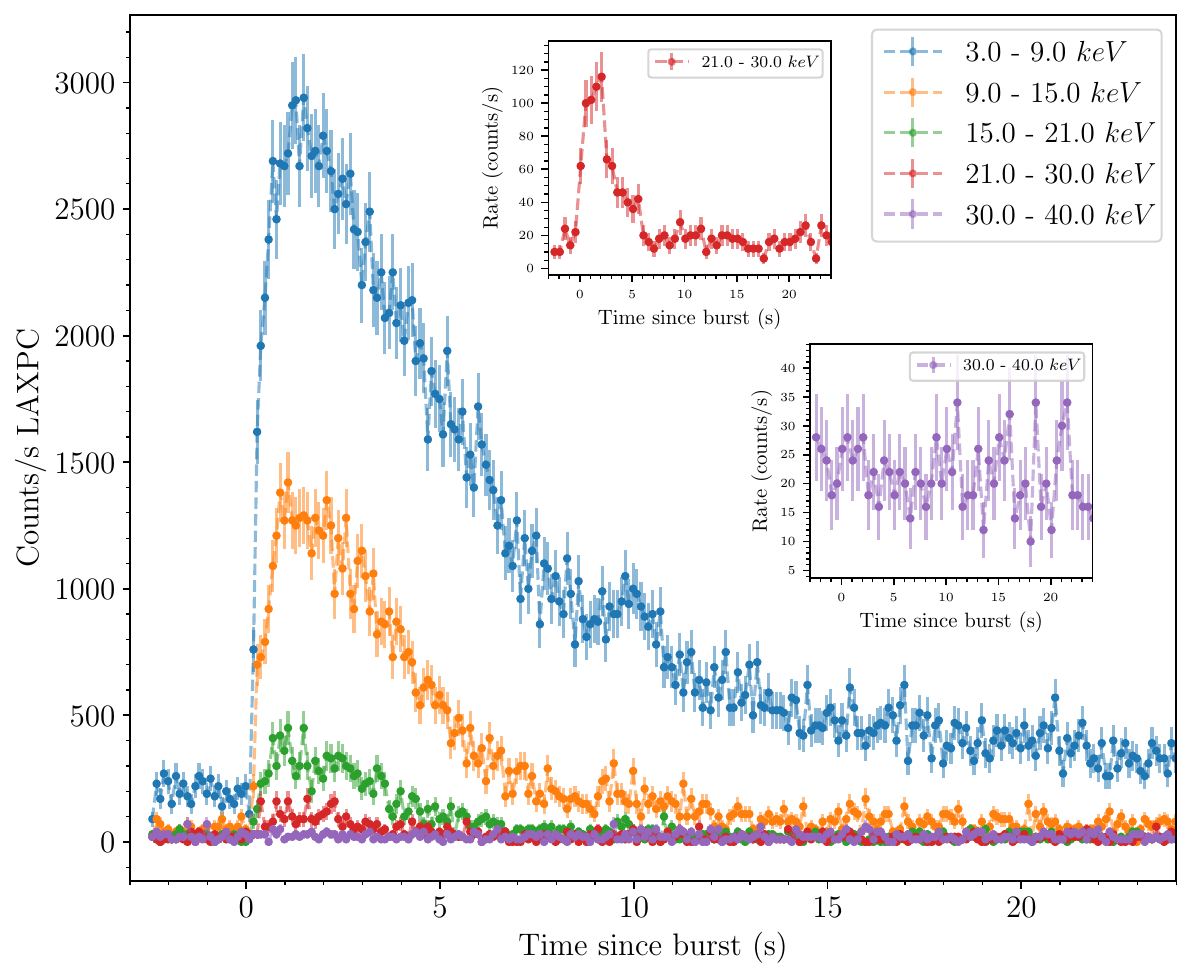}
        \caption{Same as Figure \ref{fig:energy_resolved_burst}, for B8}
        \label{fig:energy_resolved_lc_b8}
    \end{subfigure}
    \hfill
    \begin{subfigure}[b]{0.3\textwidth}
        \centering
        \includegraphics[width=\textwidth]{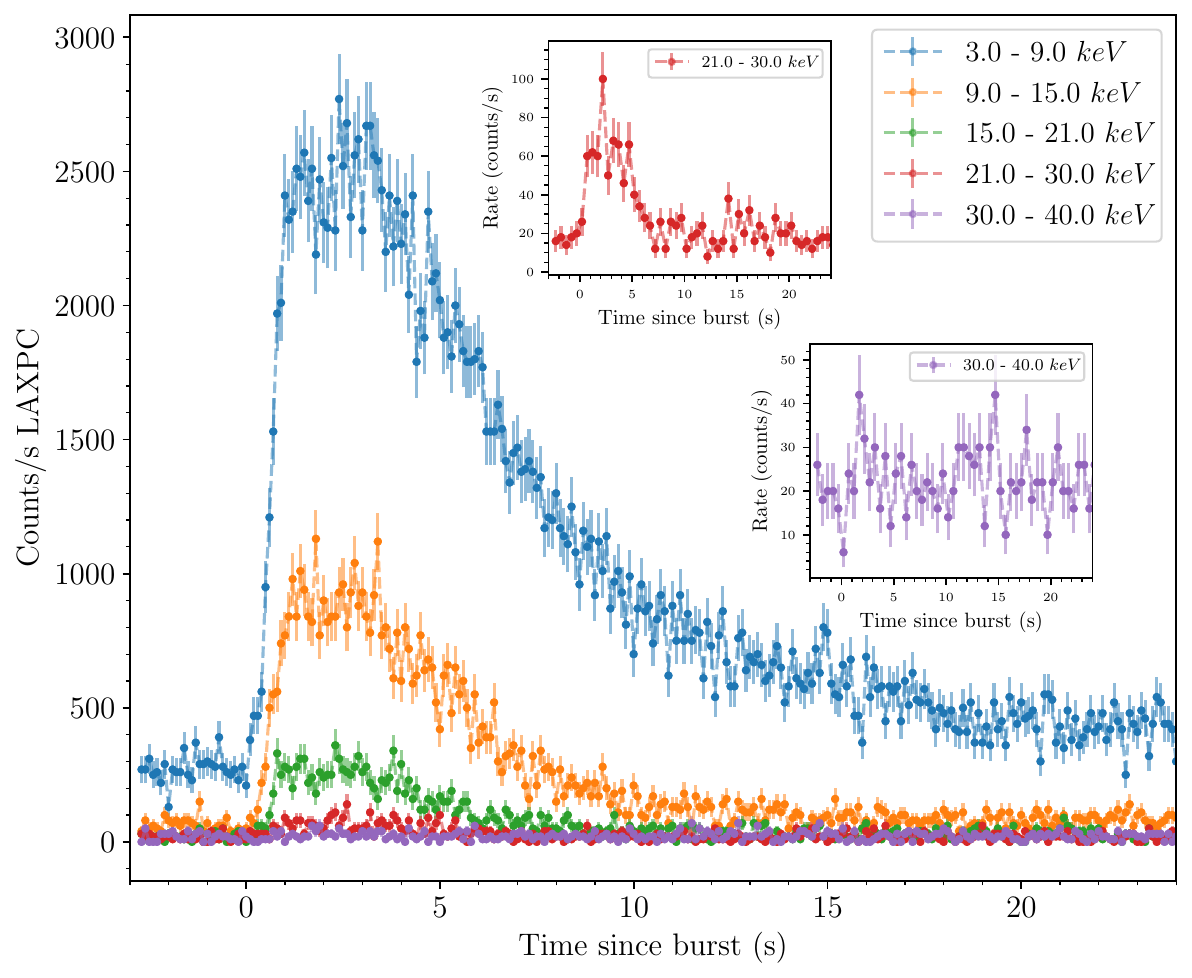}
        \caption{Same as Figure \ref{fig:energy_resolved_burst}, for B9}
        \label{fig:energy_resolved_lc_b9}
    \end{subfigure}

    \vspace{0.5cm}

    \begin{subfigure}[b]{0.3\textwidth}
        \centering
        \includegraphics[width=\textwidth]{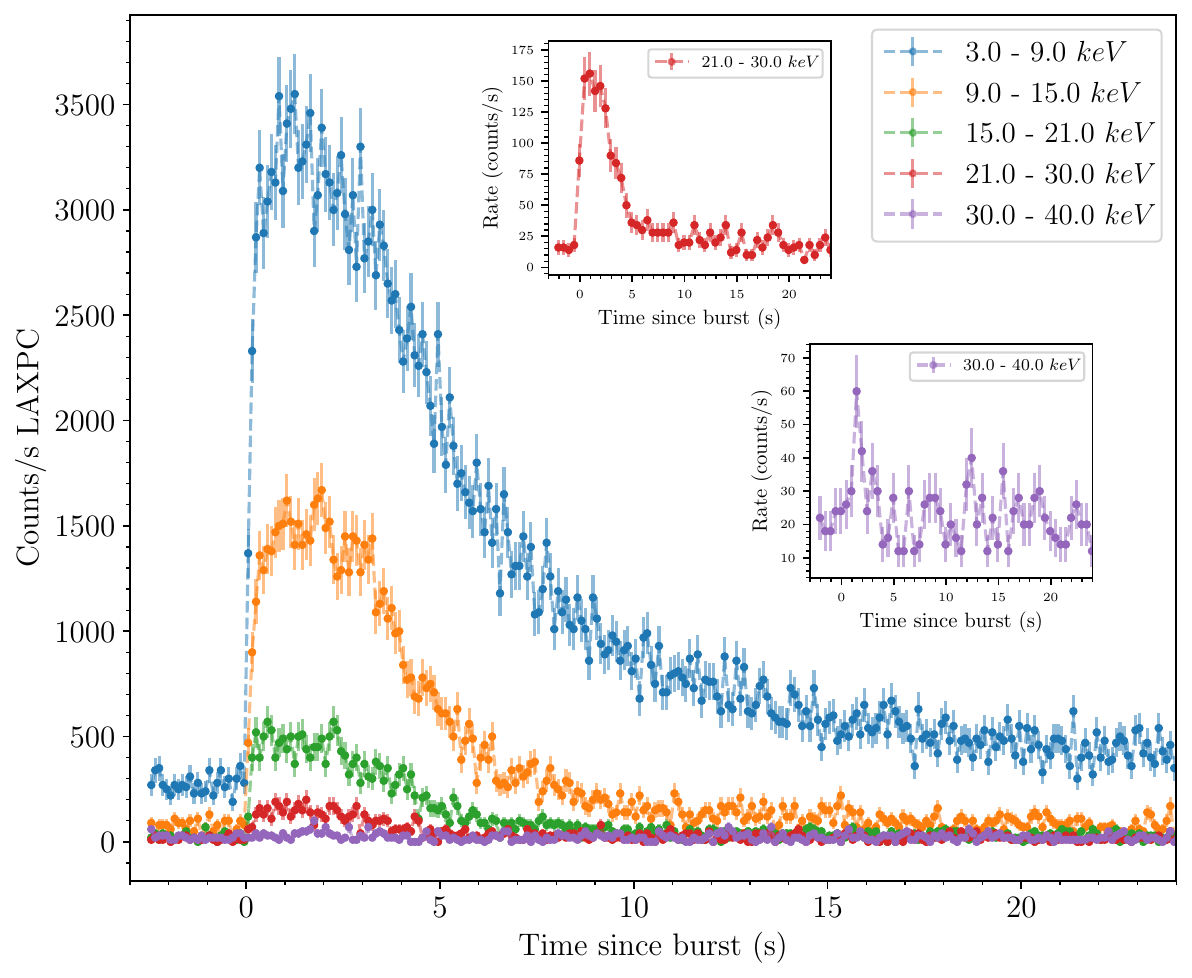}
        \caption{Same as Figure \ref{fig:energy_resolved_burst}, for B10}
        \label{fig:energy_resolved_lc_b10}
    \end{subfigure}
    \hfill
    \begin{subfigure}[b]{0.3\textwidth}
        \centering
        \includegraphics[width=\textwidth]{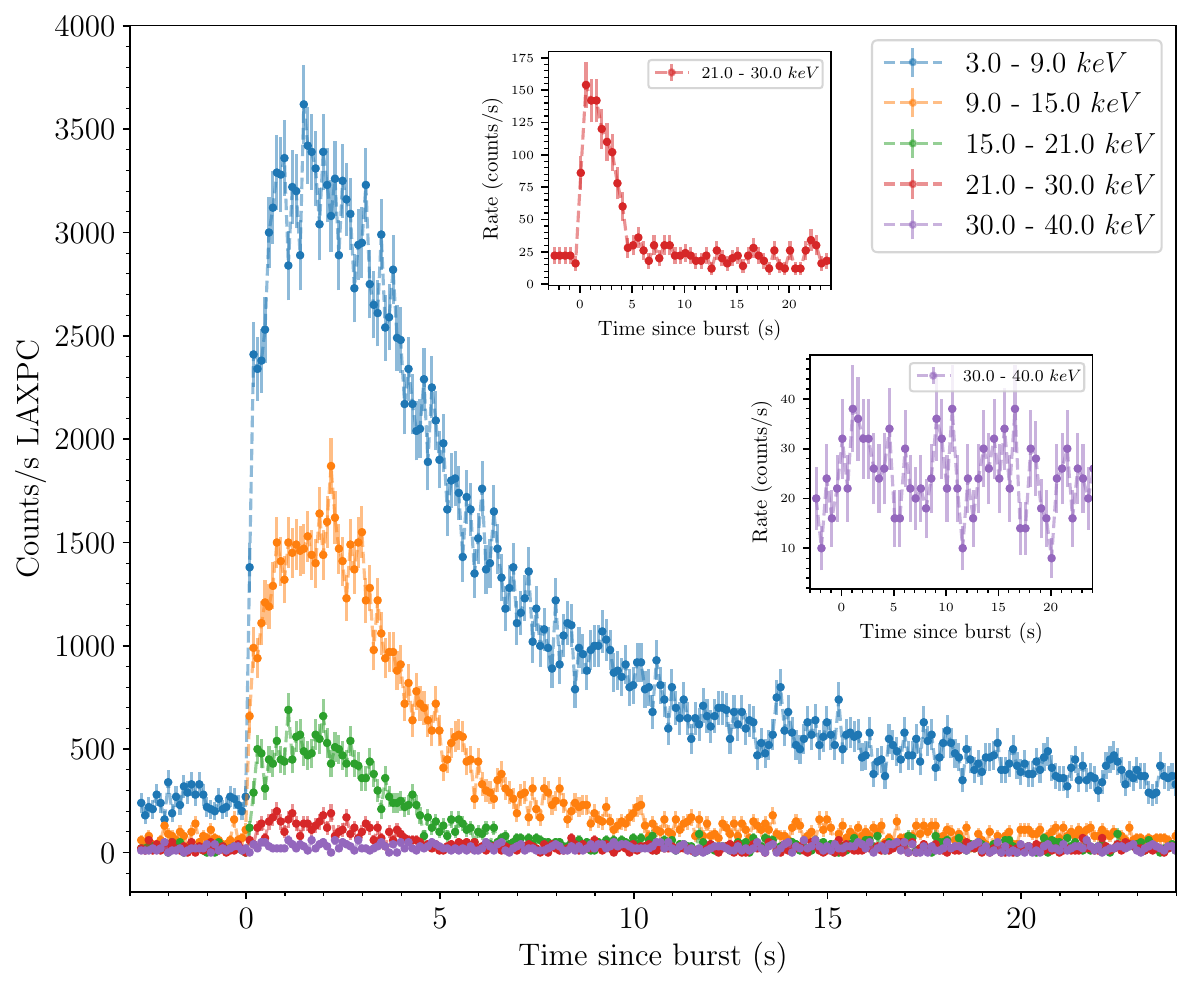}
        \caption{Same as Figure \ref{fig:energy_resolved_burst}, for B11}
        \label{fig:energy_resolved_lc_b11}
    \end{subfigure}
    \hfill
    \begin{subfigure}[b]{0.3\textwidth}
        \centering
        \includegraphics[width=\textwidth]{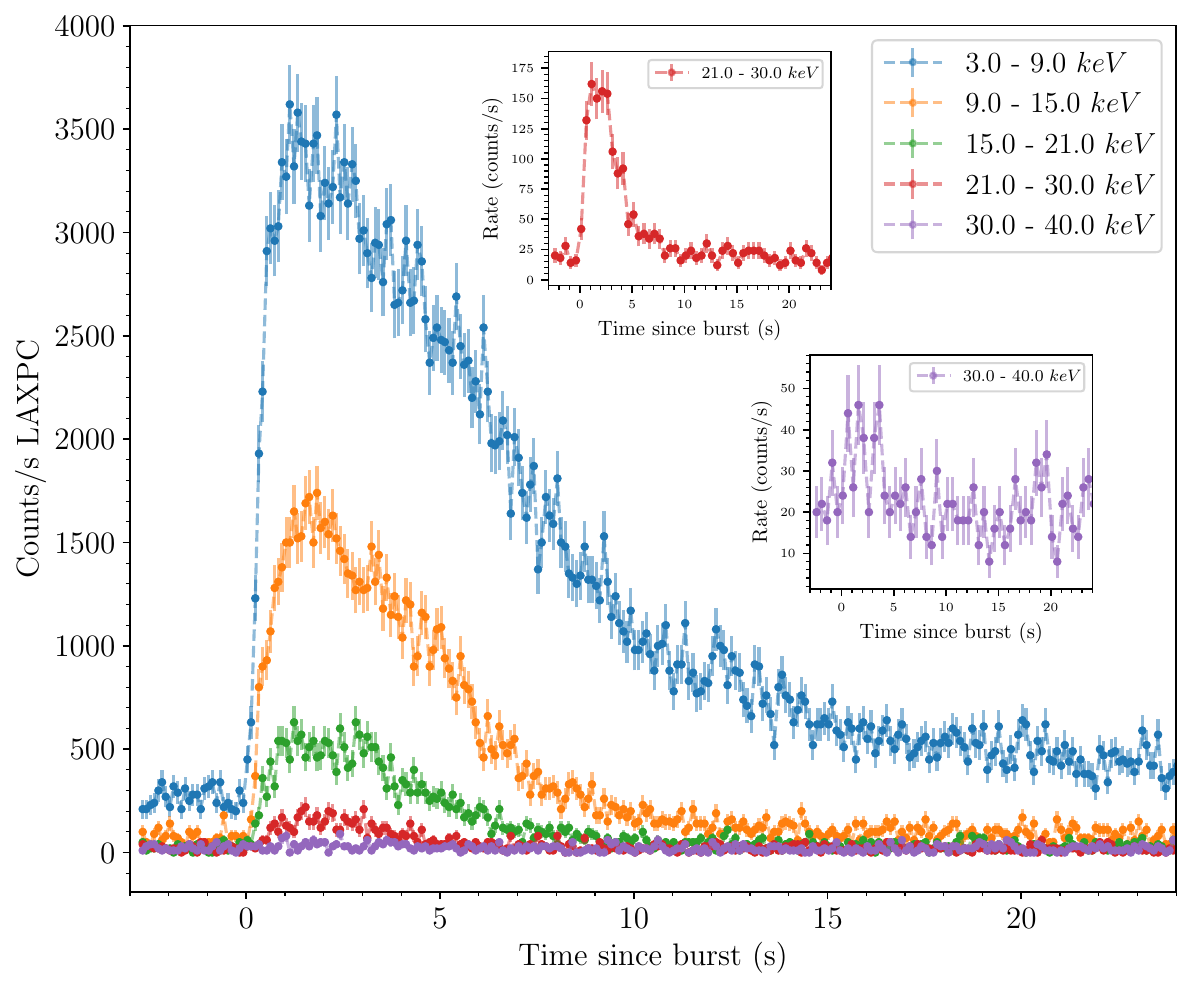}
        \caption{Same as Figure \ref{fig:energy_resolved_burst}, for B13}
        \label{fig:energy_resolved_lc_b13}
    \end{subfigure}
    
    \caption{Energy resolved burst lightcurves}
    \label{fig:all_energy_resolved_bursts}
\end{figure*}

For bursts B11, B12, B13, the burst spectrum could not be fit sufficiently with just the \texttt{bbodyrad} model\footnote{The \texttt{bbodyrad} model provides the blackbody colour temperature (T$_{bb}$) and normalization (N$_{bb}$ = (R$_{bb}$/D$_{10}$)$^2$), where R$_{bb}$ is the blackbody radius in km and D$_{10}$ is the distance to the source in units of 10~kpc.} for all the bins. We needed an additional powerlaw component to obtain a better statistically significant fit. In order to evaluate the improvement to the fit due to the addition of the powerlaw component, we calculate the F-statistic and its probability. Further, these 3 bursts showed counts $>$30.0~keV in the energy resolved lightcurves (see Figure \ref{fig:all_energy_resolved_bursts}). Details about the burst spectral parameters are presented in Table \ref{tab:burst_spec_params}.

\begin{table*}
    \centering
    \renewcommand{\arraystretch}{1.2}
    \caption{Burst oscillation candidates}
    \begin{adjustbox}{width=0.6\textwidth, center}
    \begin{tabular} {|c|c|c|c|c|}
    \hline
    \textbf{Burst} & \textbf{Frequency} & \textbf{Single trial} & \textbf{Confidence} & \textbf{Fractional}  \\
    & \textbf{(Hz)} &  \textbf{probability} & \textbf{level (\boldmath$\sigma$)} & \textbf{rms (\boldmath$\%$)} \\
    \hline
    B1 & 362.85 & $7.12 \times 10^{-7}$ & 4.06 & 3.12 $\pm$ 0.29 \\ 
    B4 & 363.73 & $3.12 \times 10^{-6}$ & 3.7 & 4.75 $\pm$ 0.47\\  
    \hline 
    \end{tabular}
    \end{adjustbox}
    \label{tab:burst_osc}
\end{table*}

\begin{figure*}[t]
    \centering
    \begin{subfigure}[b]{0.45\textwidth}
        \centering
        \includegraphics[width=\textwidth, height = 5cm]{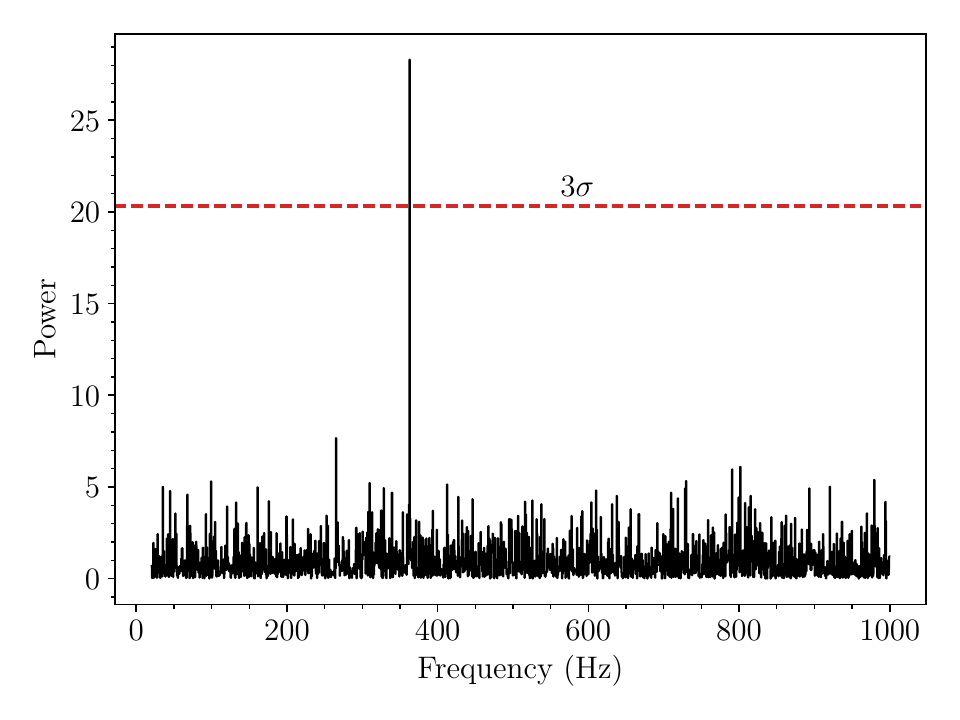}
        \caption{Power density spectrum of B1}
        \label{fig:B1-PDS}
    \end{subfigure}
    \hfill
    \begin{subfigure}[b]{0.45\textwidth}
        \centering
        \includegraphics[width=\textwidth, height = 5cm]{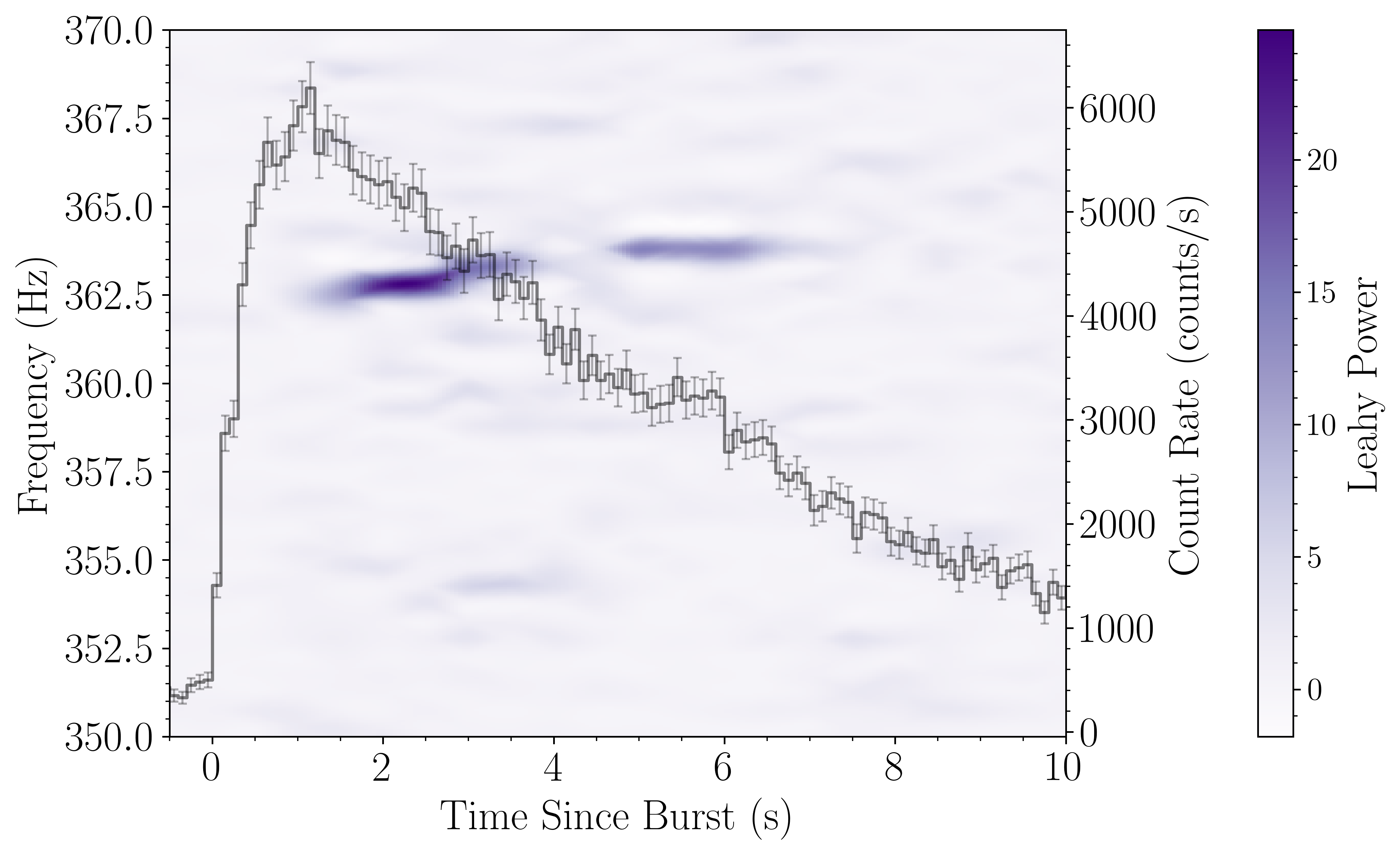}
        \caption{Dynamic power density spectrum of B1}
        \label{fig:B1-DynPDS}
    \end{subfigure}

    \begin{subfigure}[b]{0.45\textwidth}
        \centering
        \includegraphics[width=\textwidth, height = 5cm]{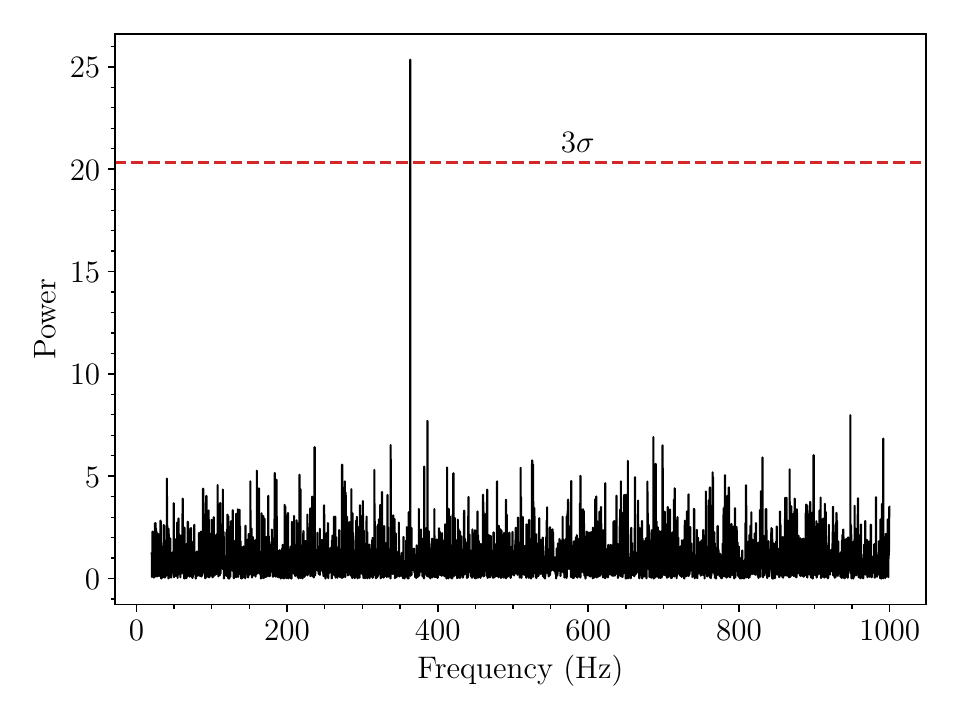}
        \caption{Power density spectrum of B4}
        \label{fig:B2-PDS}
    \end{subfigure}
    \hfill
    \begin{subfigure}[b]{0.45\textwidth}
        \centering
        \includegraphics[width=\textwidth, height = 5cm]{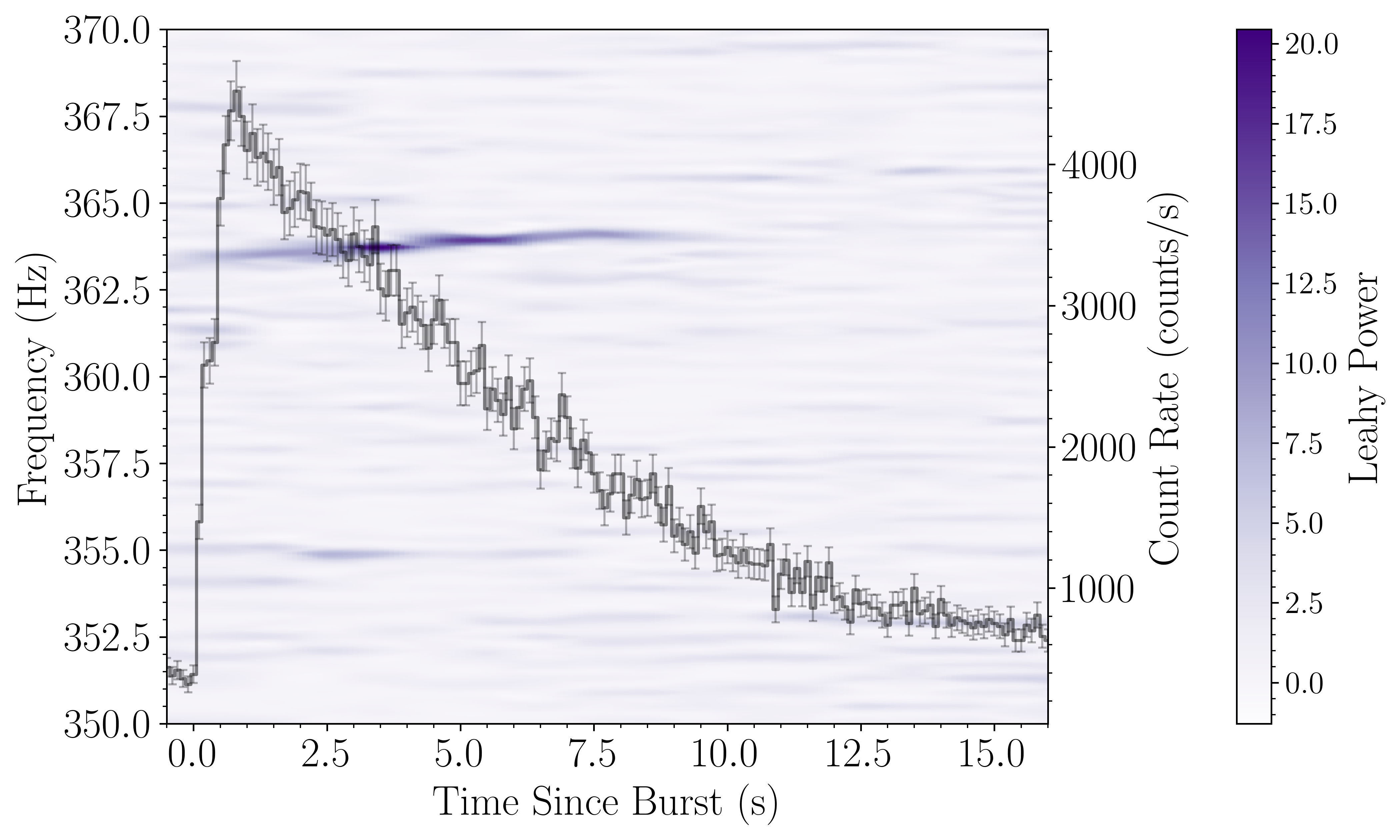}
        \caption{Dynamic power density spectrum of B4}
        \label{fig:B2-DynPDS}
    \end{subfigure}

    \caption{Left: Power density spectrum; Right: Dynamic power density spectrum. Top: Burst Oscillations of B1; Bottom: Burst Oscillations of B4.}
    \label{fig:burst_oscillations}
\end{figure*}

We calculated the bolometric flux $F_{bol}$ using the relation given by \citep{Galloway2008ApJS}
\begin{equation}
    F_{bol} = 0.001076~N_{bb}~(kT_{bb})^4~\mathrm{erg~{cm}^{-2}~s^{-1}}
\end{equation}

Since the inclination $\theta$ is not well constrained for this source, we calculate the luminosity assuming an isotropic emission using the relation $L = 4\pi R_{NS}^2 F_{bol}$. Assuming the accreted matter is evenly distributed on the surface of the NS and isotropic emission, the mass accretion rate per surface area is calculated using the below equation

\begin{equation}
\dot{M} = 6.7 \times 10^3 \, F^*_{{bol}} \, (1 + z) \, \frac{D_{10}^2}{M^*_{{NS}} \times R^*_{{NS}}}
\label{eqn:accretion_rate}
\end{equation}

Where $(1 + z) = \left( 1 - \frac{2 G M_{{NS}}}{R_{{NS}} c^2} \right)^{-1/2}$, and $z$ is the gravitational redshift due to the neutron star’s mass. $F^*_{{bol}}$ is the bolometric flux in units of $\times 10^{-9}$~erg~cm$^{-2}$~s$^{-1}$, $M^*_{NS}$ and $R^*_{NS}$ are the mass and radius of the neutron star in units of $1.4 M_{\odot}$ and $10$~km, respectively. All the calculations are presented in Table \ref{tab:burst_spec_params}.

\subsection{Search for Burst Oscillations}

We used fast Fourier Transform (FFT) to search for oscillations in 10--1000~Hz frequency range in 3.0--30.0~keV energy range for 2~s window moving forward in steps of 0.5~s from the start of the burst. The power density spectrum for all the 13 bursts were generated using \texttt{Powerspectrum} package by \texttt{Stingray} library. The obtained Leahy normalized PDS, showed local maxima at $\sim$362.85 Hz in B1, $\sim$363.66 Hz in B4. B7, B10 and B11 showed a local maxima around the BO frequency, but were sub-threshold detections. All the obtained signals are in the tail region of the bursts, which may indicate coherent Burst Oscillations (see Figure \ref{fig:burst_oscillations}).

In order to maximize the obtained power, we further extracted PDS in a 4~second window with varying segment sizes of 1~s, 2~s and 3~s in a step of 0.1~s which accounts for 40+20+10 = 70 overlapping segments, which in our case are considered as the total number of trials. The single trial chance probabilities and their significance have been calculated using the relation $x = e^{-P_{max}/2}~\times~n$, where $P_{max}$ is the maximum Leahy power measured and $n$ is the number of trials over which the detection was made and $X = \sqrt{2} erf^{-1} (1 - x)$ where X$\sigma$ corresponds to detection significance. (see e.g. \cite{Roy2021MNRAS}). Further, the fractional rms amplitudes are calculated using the relation:

\begin{equation}
    A_{\text{rms}} = \left( \frac{P_s}{N_m} \right)^{1/2} \times \frac{N_m}{N_m - N_{\text{bkg}}}
    \label{eqn:rms}
\end{equation}

where $P_s$ is the power of the signal, $N_m$ and $N_{bkg}$ are total and background count rates during the interval searched. An approximate relation $A_{\mathrm{rms}}$ = $\sqrt{P_s/N_m}$ can be used, as N$_{bkg}$ is negligible compared to high count rate ($N_m$) during the burst. The details of the burst oscillation candidates are presented in Table \ref{tab:burst_osc}. 

\section{Discussion}\label{sec:discussion}
\source is one of the most widely studied objects among the Atoll population. In this work, we analyse some additional properties of this system, using detailed timing and spectroscopic methods to better characterize this class of objects, as a whole. By correlating the burst properties, presence/absence of QPOs, and hardness diagrams, we derive several insights into the physical geometry of accretion powered systems.

\subsection{PRE bursts, touchdown fluxes and non-thermal burst emission}\label{sec:PREdisc}

We adapt the method described by \citet{Guver2012ApJ} to identify the PRE bursts.
Using the PRE bursts identified as part of this analysis, (B1, B2, B3, B6, and B8), we further infer the source distance (as described by \citealt{Basinska1984ApJ}), Eddington limits and compare them with previous estimates. Assuming a canonical NS radius of 10~km  and a mass of 1.4~M$\odot$, we obtain a distance estimate of 5.18--5.21~kpc which is consistent with observations from \textit{RXTE} \citep{Galloway2008ApJS} and \textit{NICER} \citep{Bostanci2023ApJ}. We further estimate the Eddington limit using the peak burst flux to be $\sim$1.46~$\times~10^{38}$~erg~s$^{-1}$. Using the bolometric flux obtained from the analysis of the persistent spectral fitting in Section \ref{sec:per-spec}, we infer that the source was accreting at 13.6\% Eddington (for B1), 4.9\% Eddington (for B2 and B3), 9.4\% Eddington (for B6 and B8). The average bolometric flux $F_{peak,bol}$, during the peak of the bursts is found to be $\sim$4.6~$\times~10^{-8}$~erg~cm$^{-2}$~s$^{-1}$, which is smaller compared to the average peak flux of $9.4\pm3.6$~$\times~10^{-8}$~erg~cm$^{-2}$~s$^{-1}$ in MINBAR \citep{Galloway2020}.

The continuum emission from \tnbs has been conventionally described using a simple blackbody, although several cases have observed deviations from a purely thermal spectrum \citep{Bhattacharya2018ApJ}. For example, the scattering of the burst photons in the NS atmosphere can lead to the hardening of the spectrum \citep{Bhattacharyya2010AdSpR, Galloway2021ASSL}. The requirement of the non-thermal powerlaw component in three of the bursts, and the presence of hard photons ($>$20~keV), in this work, is presumably a manifestation of such a comptonized effect. We note that a substantial subset of our bursts are observed in the high soft state of the source. A similar non-thermal component addition has been previously reported for 4U 1608--62 \citep{Guver2021ApJ}, although, in that case, the source was in the low-hard state.

Thermonuclear burst photons can in some cases lead to an increased accretion rate, which could increase the observed persistent emission \citep{Bhattacharya2018ApJ, Worpel2015ApJ}. Alternatively, enhancements in the persistent emission near the burst can occur due to burst reprocessing in the accretion disk \citep{intZand2013}. Whichever may be the case, it is becoming increasingly evident that the burst emission affects the accretion process and could lead to variability in the persistent flux as measured during the burst. To deal with such a possibility, a new method called the ``variable persistent flux method” ($f_a$-method, \citealt{Worpel2013, Worpel2015ApJ}) has been developed. However, this method is a favourable fitting option for instruments sensitive to energies below 2~keV as it accounts for variability in the persistent emission during \tnbs, often influenced by reprocessing effects prominent in this energy range. Additionally, for the cases of 4U 1608--52, and XTE J1739--285, which have a large line of sight $N_H$ absorption component, the use of scaling factor has not yielded any significant differences in the fitting process \citep{Guver2021ApJ, Guver2022MNRAS, Bult2021ApJ}.


\subsection{QPOs, Burst Oscillations and source state evolution}
Some of the fastest timing signatures that are detected from accreting NS systems are the kHz QPOs ($\sim$100~$\mu$s), which therefore serve as a direct probe of innermost accretion flows \citep{van2000}. The power spectral analysis of the burst-free light curve resulted in the detection of twin kHz QPOs at $619\pm10$~Hz and $965\pm6$~Hz. Such a simultaneous detection of an upper and lower kHz QPO has been previously reported during several observations of \source in the past (see for example., \citealt{DiSalvo2001ApJ, Mendez2001ApJ}). In addition, a series of QPO triplets (upper kHz, lower kHz and low frequency), detected using \asat , have been recently modelled using the Relativistic Precession Model (RPM) to constrain the mass and moment of inertia of the NS \citep{Anand2024ApJ}. Their work showed that, in general, the QPO measurements favoured stiffer Equations of State (EoS). 

\begin{figure}
    \centering
    \includegraphics[width=\columnwidth,height=6cm]{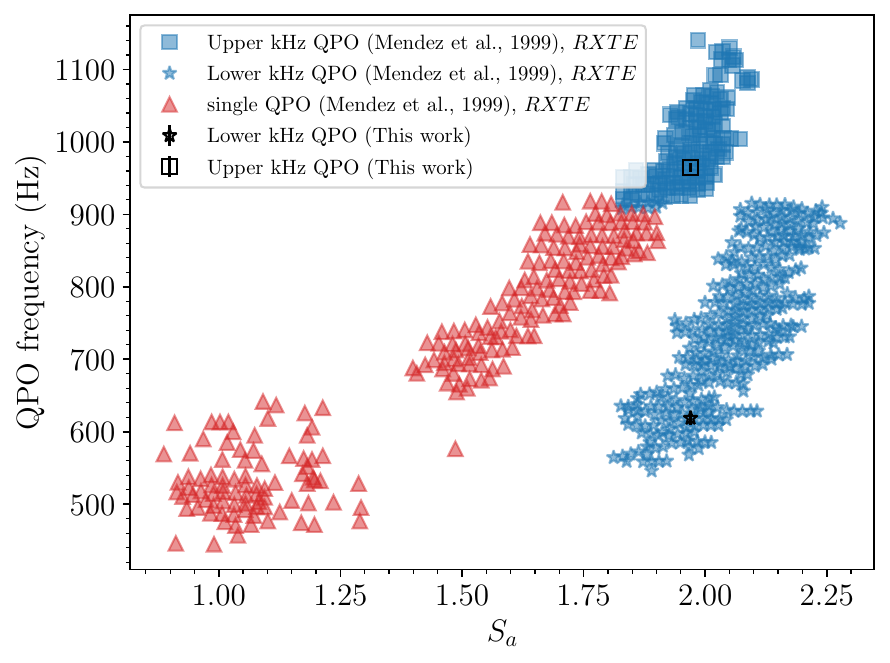}
    \caption{Frequencies of the upper and lower kHz QPOs plotted as a function of source position on the Hardness Intensity Diagram ($S_a$). Blue squares and * points represent lower and upper kHz QPOs respectively, and red data points are where only one QPO was observed. This figure is adapted from \citep{1999ApJ...517L..51M}.}
    \label{fig:qpo_sa}
\end{figure}

Typically, the upper kilohertz QPO frequency ($\nu_u$) is assumed to be the Keplerian orbital frequency ($\nu_K$) of the accretion plasma \citep{Wang2018}. It can be used to estimate the emission radius of the kHz QPOs, which can be considered the same as the magnetospheric disk radius \citep{Lewin2006csxs}.

\begin{equation}
\nu_{ukhz} = \sqrt{\frac{GM}{4 \pi^2 r^3}}
\label{eqn:ukhz}
\end{equation}

where $G$ is the Gravitational constant, $M$ is the mass of the NS and $r$ is the emission radius of the kHz QPO referring to the NS centre i.e. the magnetospheric disk radius.

The co-rotation radius $r_{co}$ in NS-LMXB is the radial distance referring to the NS centre where the accretion plasma co-rotates with the NS magnetosphere, meaning that the Keplerian orbital frequency $\nu_K$ there equals the NS spin frequency $\nu_s$ \citep{1991PhR...203....1B}. Setting $\nu_K$ = $\nu_s$ in Equation \ref{eqn:ukhz} gives the corotation radius $r_{co}$ as

\begin{equation}
    r_{co} = \frac{GM}{4 \pi^2}^{1/3} \nu_{s}^{-2/3}
    \label{eqn:corotation_radius}
\end{equation}

With the upper kHz QPO detected at $965\pm6$~Hz, we obtain the co-rotation radius of $17.2\pm0.06$~km. Assuming the dipolar magnetic field of NS and upper kHz QPO originated at the Alfven radius, the magnetic field strength $B_{r}$ at the magnetospheric disk radius `$r$' is calculated as:

\begin{equation}
    B_{r} = B_s (R_{NS}/r_{m})^3
    \label{eqn:magnetic_field}
\end{equation}

Assuming $B_s$ $\sim$ 1.8--6.5~$\times~10^8$~G \citep{Mondal2017MNRAS}, $B_{r}$ is calculated to be 0.35--1.27~$\times~10^7$~G. 




We compile a list of all previously detected high-frequency kHz QPOs using RXTE \citep{Mendez2001ApJ} and \textit{AstroSat}/LAXPC (this work) and indicate the peak frequencies as a function of the source spectral hardness $S_a$ (see Figure \ref{fig:qpo_sa}). The twin kHz phenomenon is a consequence of the source being in the `banana' state. The lower frequency further gets suppressed at lower accretion rates, consistent with non-detections in the other observations analysed in our work. The absence of the upper kHz QPO frequency in lower accretion states observed with LAXPC, which was consistently detected in past \textit{RXTE} observations, will be explored in future studies. The separation between the twin kHz QPO frequencies is typically found to be correlated in some form or the other to the NS spin frequency (for example, see \citealt{Miller1998ApJ,Lamb2001ApJ,Wijnands2003Natur,Lee2004ApJ,Lewin2006csxs}). The more commonly invoked beat frequency model (a.k.a. sonic-point model; \citealt{Miller1998ApJ}) suggests that the lower kHz QPO is observed at the beat frequency between the NS spin and the upper QPO due to the interaction of the neutron star’s magnetic field with matter at the inner edge of the disk. 

BOs are established to be rotationally induced modulations that occur due to temperature variations on the NS surface (see \citealt{AnnaWatts2012} for a detailed review). The detection of a BO very close to the NS spin in the pulsar source SAX J1810.8--2609, firmly established the BO-spin frequency correlation \citep{Wijnands1998Nature}. The $\sim$363 Hz BO frequency of \source has remained steady over the years and is consistent with our timing search results for bursts B1 and B4 bursts.

During the course of the seven \asat observations considered in this paper, \source has evolved through several spectral states. The observations are separated by several years, O1 being the earliest (2016), while the remaining observations are from a later epoch, i.e., 2018 and 2019. As observed in almost all the Atoll sources, \source also exhibits a variety of spectral and timing signatures during these epochs. Interestingly, the results from O1 stand out from the rest. During this particular observation period, the source is found to be in an extremely soft ``banana” state, corresponding to a spectral powerlaw index of 2. Twin kHz QPOs are strongly detected in the power spectrum, with a peak-to-peak separation of around 346~Hz which is slightly lesser than the Burst Oscillations at 363~Hz found in the  cooling tail of B1 (see Figure \ref{fig:B1-PDS}), which belongs to the same observation. While the QPOs are found only during the start of the observation, the burst in the same observation is found much later. The simultaneous detection of kHz QPO and BO, allows us to further strengthen the association between the separation frequency, the BO and the NS fundamental spin frequency. 

An important consequence of spectral state changes is that the variations in the accretion rate allows for sampling of the magnetospheric radius at various radii. This can be effectively probed using kHz QPOs as has been demonstrated by \citet{Wang2018}. A compilation of all the previous \textit{RXTE} observations in various spectral states for \source shows that the high/soft `banana state' is where the field measurements are relatively stronger and the associated high frequency phenomena occur closer to the NS surface. This particular observation epoch (O1) samples a magnetic field strength of 2~$\times$~10$^{7}$~G at a magnetospheric disk radius of 17~km. 

Since we have a simultaneous detection of twin kHz QPOs and BO from the same observations, we have two independent handles for correlating these frequencies with the NS spin period. We indicate all the previous measurements of the difference between the twin kHz QPO frequencies ($\nu_{sep}$) against the lower kHz QPO frequency ($\nu_l$). In order to be consistent with the energy ranges, we compare all the \textit{RXTE} and \textit{AstroSat} measurements from \citet{Mendez2001ApJ} and \citet{Anand2024ApJ}, respectively, with our observation now included. We plot the band of BO frequency values for comparison. We observe several interesting effects in Figure \ref{fig:delta_qpo}. While the BO frequencies occupy a very narrow range, the range of $\nu_{sep}$ has a lot more spread. Particularly, for the case of O1 which shows both the BO and the twin kHz QPOs, we observe that the separation frequency is slightly lower than the BO frequency, but is consistent within uncertainties. This allows us to better constrain kHz QPO correlation models. Our current observations strongly prefer the sonic-point beat frequency model, which interprets the difference to be close to the NS spin frequency. In addition to this, we observe a decreasing trend of $\nu_{sep}$ at larger $\nu_l$. Such trends may hint towards an underlying physics that will be useful to better understand the origin of kHz QPOs.

\subsection{Conclusions}
We present a comprehensive study of the Atoll Source \source using archival \asat observations. By performing detailed spectro-temporal studies of persistent emission and \tnbs, we have been able to derive several key insights into the physical properties of the source.

\begin{itemize}
\item From the Hardness Intensity Diagram, we observe the source to be in multiple intensity states. A total of thirteen \tnbs have been observed, with one burst in O1 and five bursts in O7, which belong to high intensity states. Five bursts were observed in O6 and two bursts in O5, which are intermediate and low intensity states respectively.
\item O1 shows the presence of twin kHz QPOs, with a peak-to-peak separation of $\sim$ 346 Hz, which is slightly smaller than the frequency of the reported Burst Oscillation (BO) at $\sim$ 363 Hz. The presence of kHz QPOs and BO in the same observation allows us to better constrain the kHz QPO correlation models.
\item The fractional rms of the upper kHz QPO increases with energy, and reaches a maximum amplitude of $\sim$ 6 \% around 16 keV. This indicates that the 16 keV photons are likely causing upper kHz QPO.
\item From the upper kHz QPO frequency, we infer the magnetospheric disk radius to be $\sim$ 17.2 km, and the magnetic field strength at the magnetospheric disk radius to be in the range of 0.35 - 1.27 $\times 10^7$ G.
\item The persistent spectra could be well described by a simple blackbody model, with the addition of a powerlaw component. The powerlaw spectral index $\Gamma$ is found to be around $\sim$ 2 for all the observations. No disk reflection features were observed.
\item From the energy resolved burst lightcurves, we see that all of the \tnbs show the presence of photons $>$ 20 keV. Some bursts show photons $>$ 30 keV. The presence of hard photons in the burst emission is indicative of a non-thermal component in the burst emission.
\item We infer that the source was accreting at 13.6 \%, 4.9\%, 9.4\% and 13.3\% of the Eddington limit during O1, O4, O6 and O7 respectively, where \tnbs are present. This suggests that the bursts are dominated by helium ignition after steady hydrogen burning via the CNO cycle.
\item Among the 13 bursts, 5 bursts (B1, B2, B3, B6, and B8) were identified as PRE bursts. The touchdown radius is found to be in the range of 3.18--5.54 km. The source distance was estimated from the peak Luminosities to be 5.18--5.21 kpc. These measurements are consistent with previous observations.
\item Coherent Burst Oscillations were detected in B1 and B4 at 362.85 Hz and 363.73 Hz respectively in the cooling tail of the bursts. The fractional rms amplitude of the Burst Oscillations was found to be 3.12 $\pm$ 0.29 \% and 4.75 $\pm$ 0.47 \% respectively.

\end{itemize}

\section{Acknowledgements}
We thank the Indian Space Research Organisation (ISRO) for financial support for the project ``Timing and spectral studies of Type-1 thermonuclear X-ray bursts using AstroSat.'' (DS 2B-13013(2)/3/2020-Sec.2 dtd. 12.01.2021). This work utilises data from LAXPC and SXT payloads onboard \asat made available through the Indian Space Science Data Center (ISSDC). 

\bibliographystyle{apj}
\bibliography{references}

\section{Appendix}

\begin{figure*}[!htbp]
    \centering
    \vspace{-10pt}
    \begin{subfigure}[b]{0.3\textwidth}
        \centering
        \includegraphics[width=\textwidth]{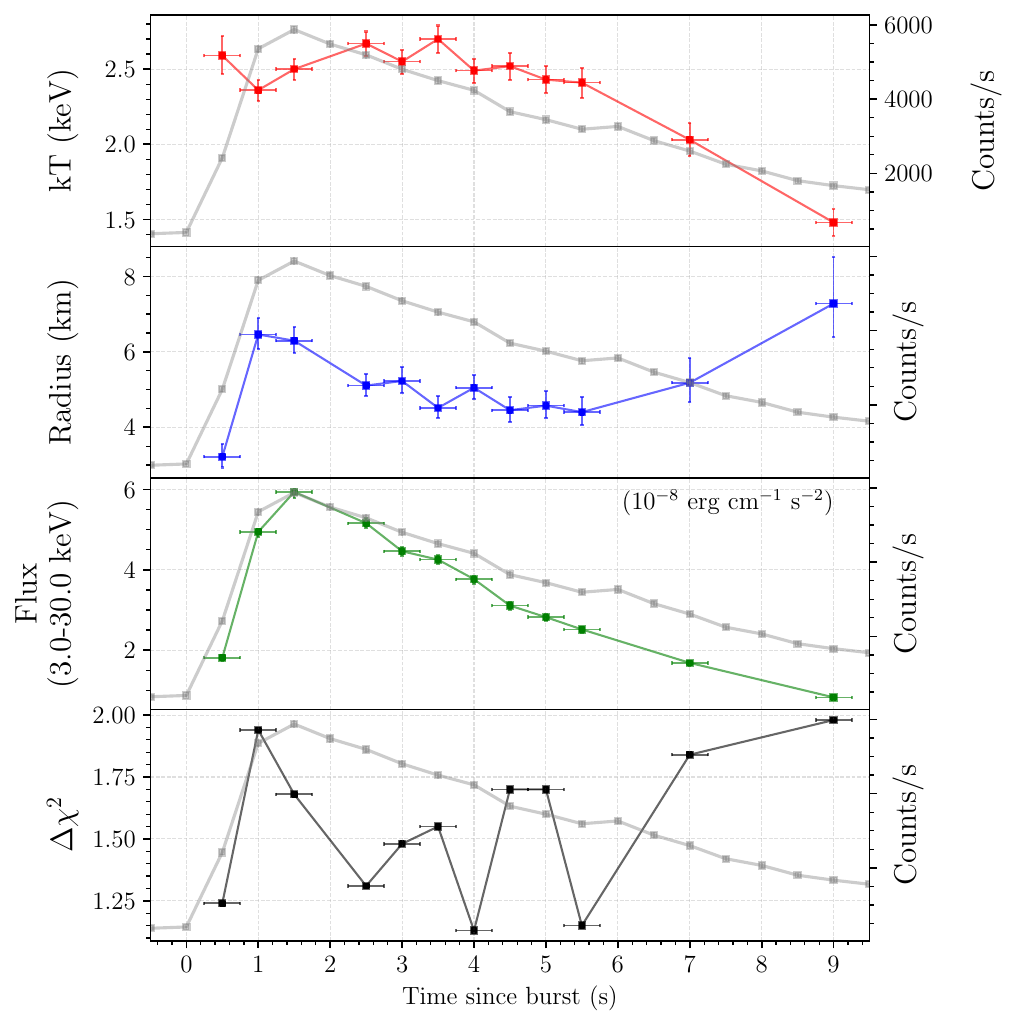}
        \caption{Same as Figure \ref{fig:time_resolved}, for B1}
        \label{fig:0578_burst_1}
    \end{subfigure}
    \hfill
    \begin{subfigure}[b]{0.3\textwidth}
        \centering
        \includegraphics[width=\textwidth]{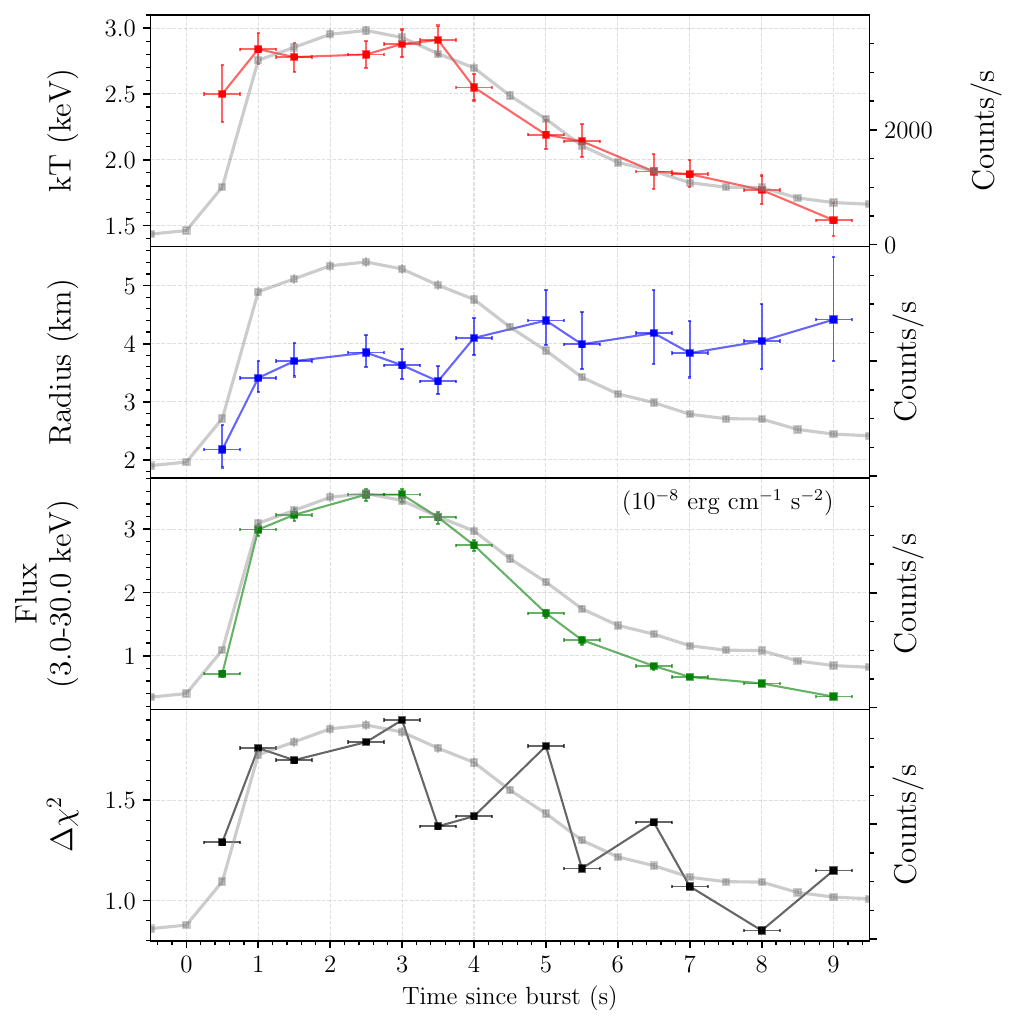}
        \caption{Same as Figure \ref{fig:time_resolved}, for B2}
        \label{fig:2254_burst_1}
    \end{subfigure}
    \hfill
    \begin{subfigure}[b]{0.3\textwidth}
        \centering
        \includegraphics[width=\textwidth]{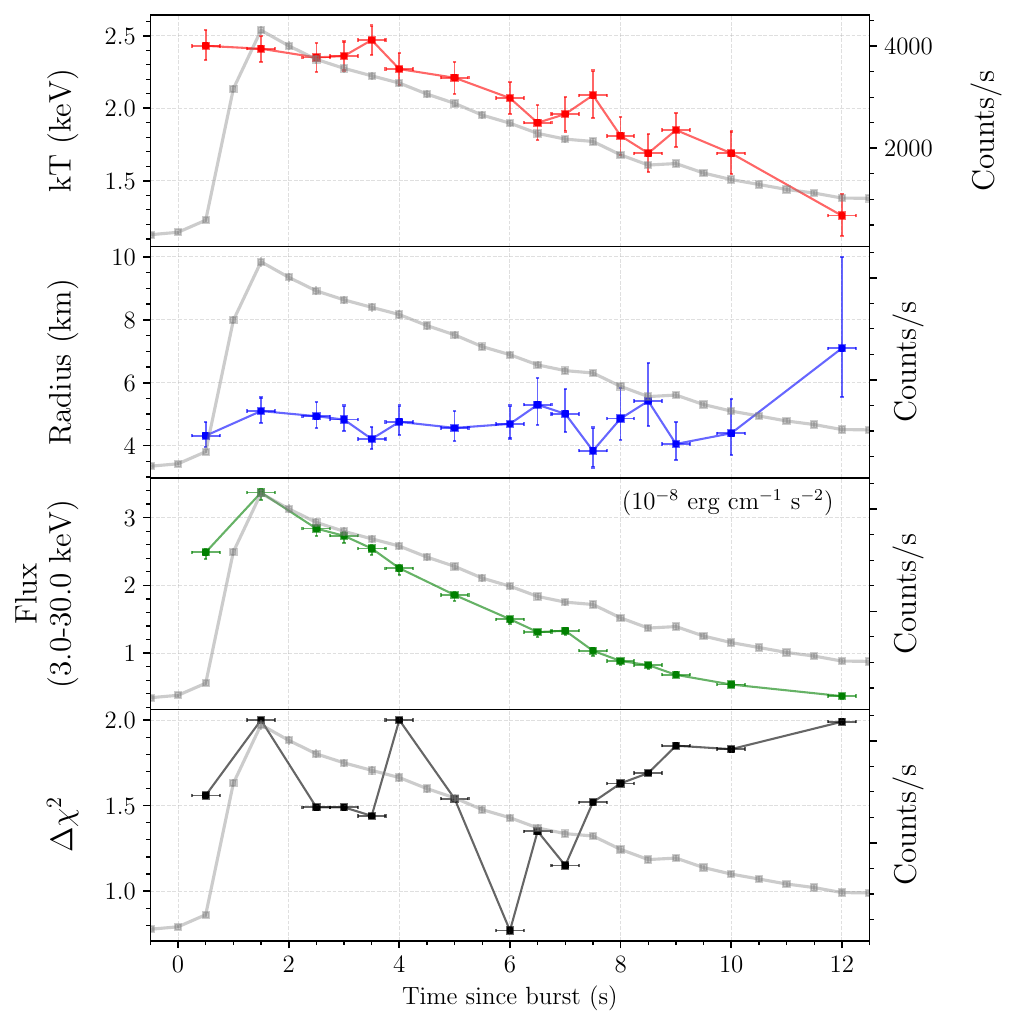}
        \caption{Same as Figure \ref{fig:time_resolved}, for B4}
        \label{fig:2890_burst_1}
    \end{subfigure}
    
    \begin{subfigure}[b]{0.3\textwidth}
        \centering
        \includegraphics[width=\textwidth]{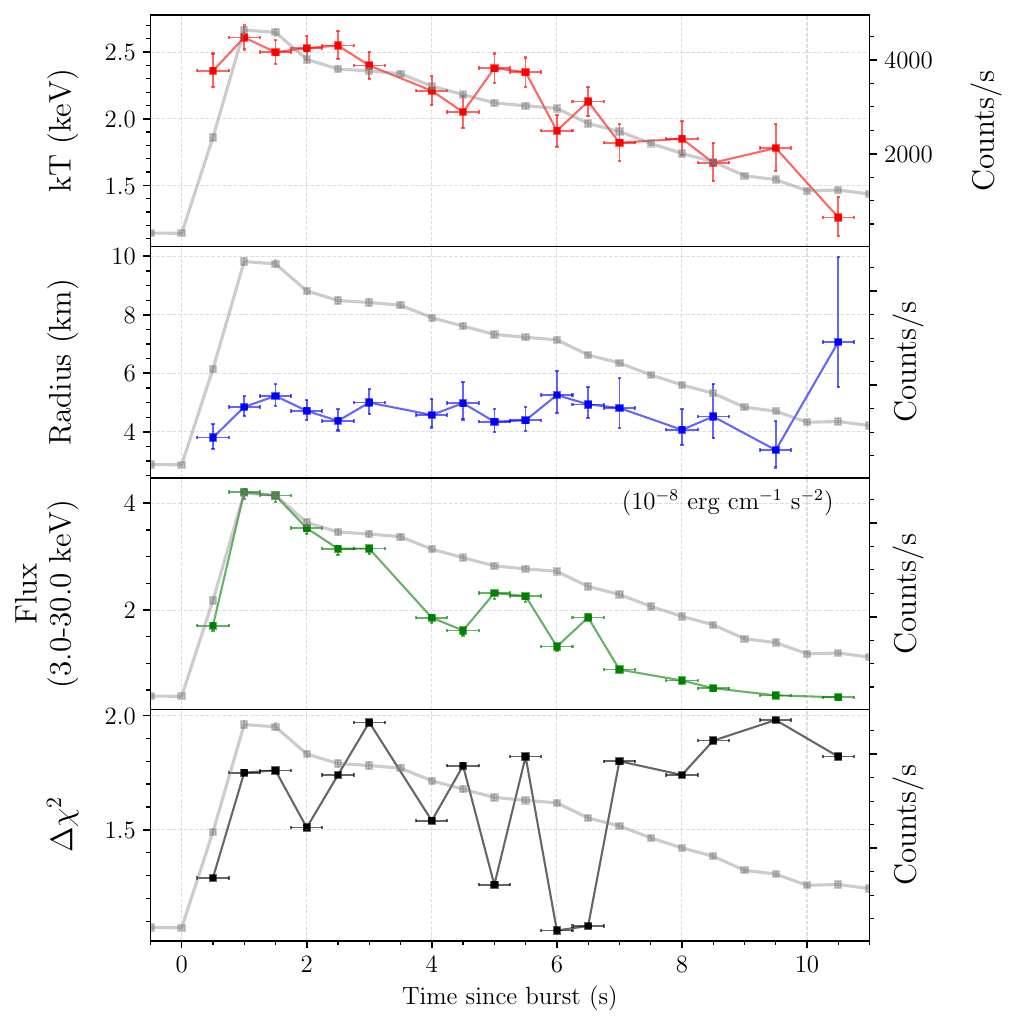}
        \caption{Same as Figure \ref{fig:time_resolved}, for B5}
        \label{fig:2890_burst_2}
    \end{subfigure}
    \hfill
    \begin{subfigure}[b]{0.3\textwidth}
        \centering
        \includegraphics[width=\textwidth]{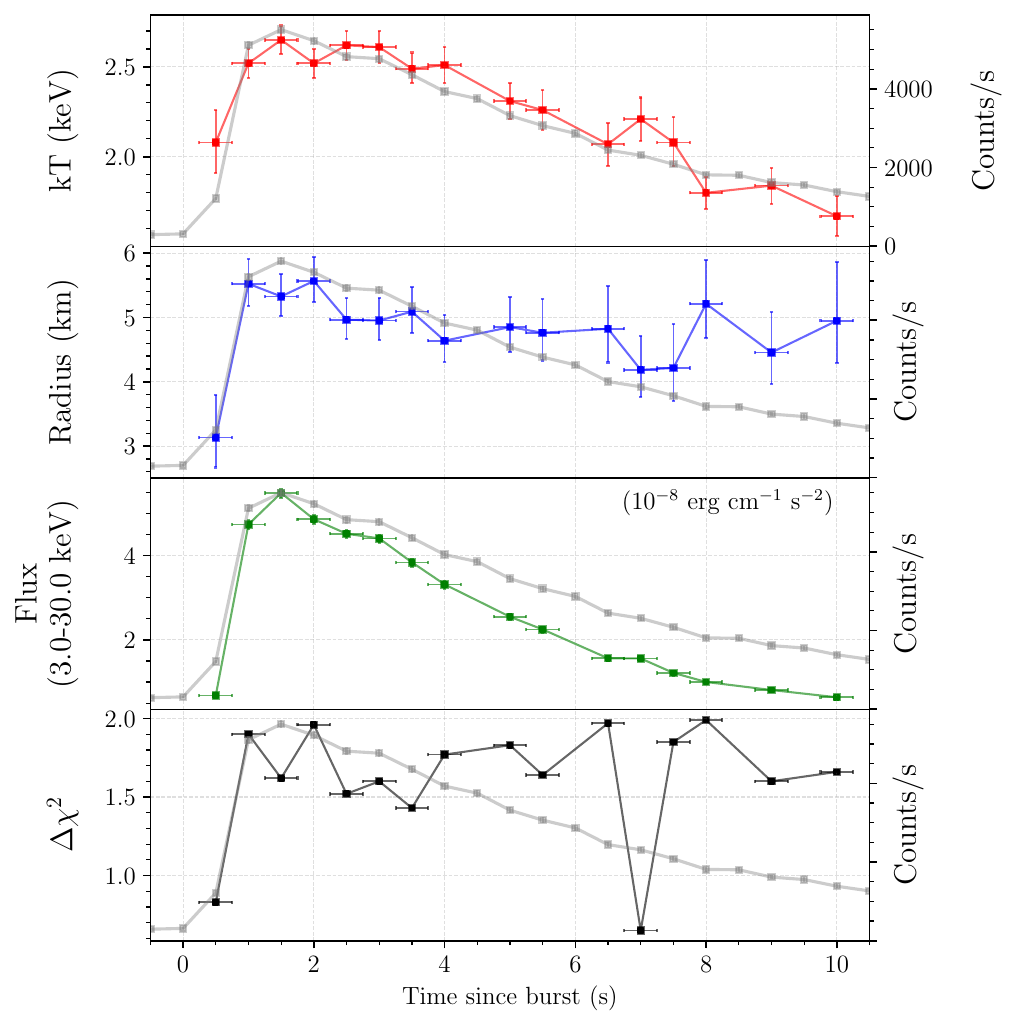}
        \caption{Same as Figure \ref{fig:time_resolved}, for B6}
        \label{fig:2890_burst_3}
    \end{subfigure}
    \hfill
    \begin{subfigure}[b]{0.3\textwidth}
        \centering
        \includegraphics[width=\textwidth]{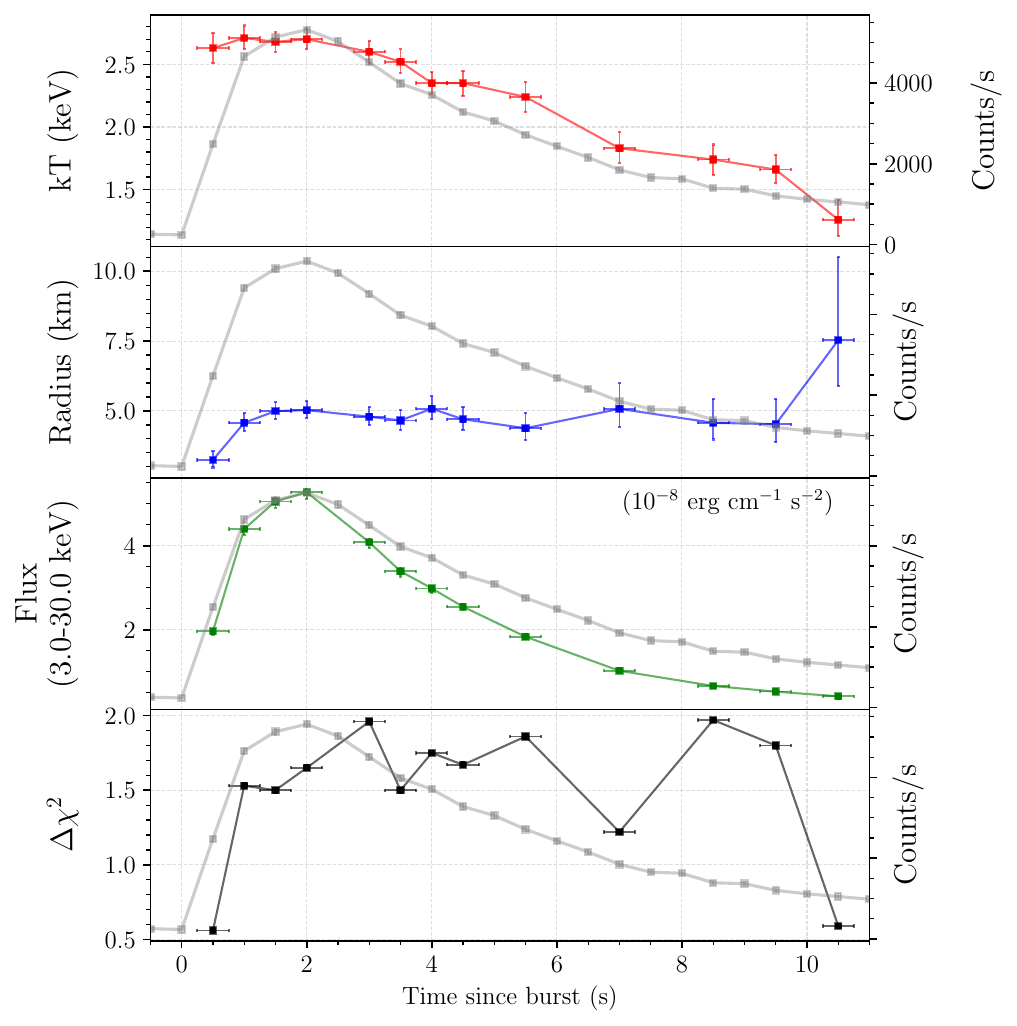}
        \caption{Same as Figure \ref{fig:time_resolved}, for B7}
        \label{fig:2890_burst_4}
    \end{subfigure}
    
    \begin{subfigure}[b]{0.3\textwidth}
        \centering
        \includegraphics[width=\textwidth]{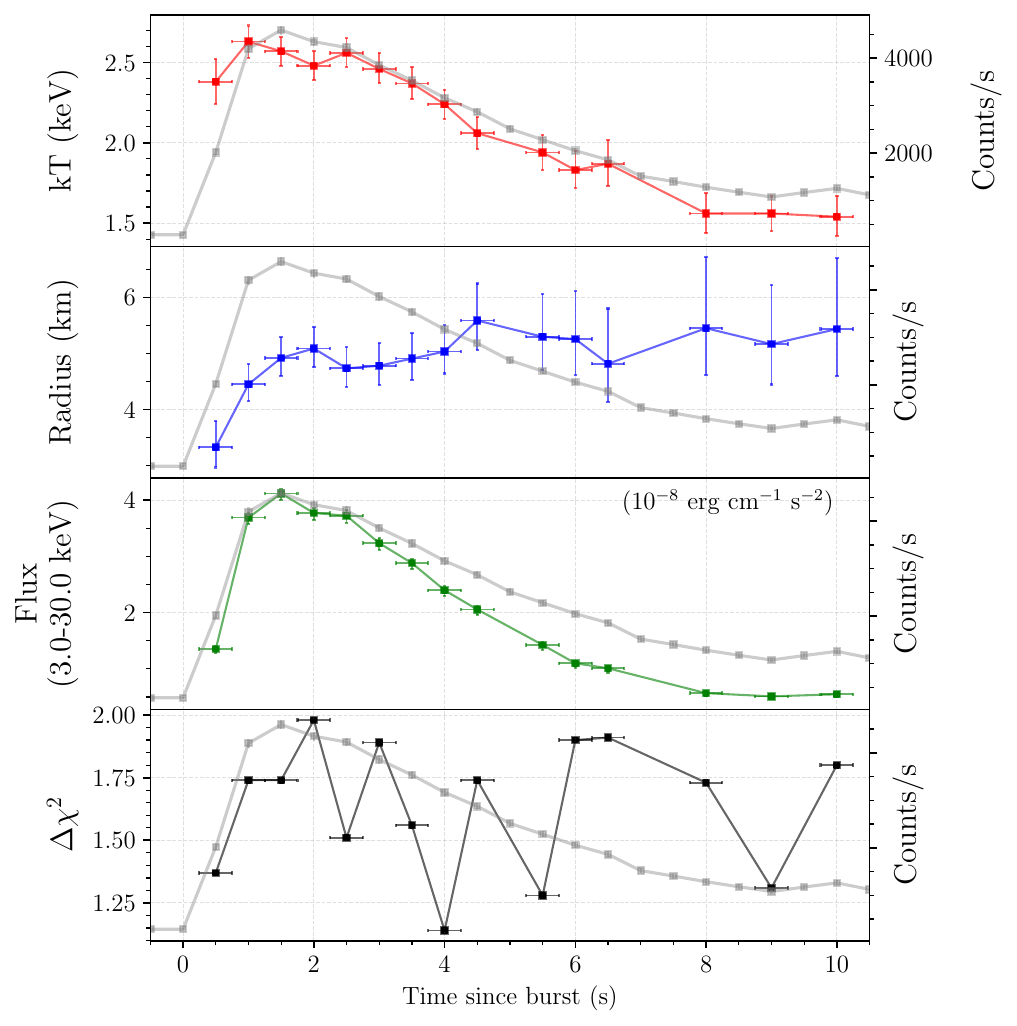}
        \caption{Same as Figure \ref{fig:time_resolved}, for B8}
        \label{fig:2890_burst_5}
    \end{subfigure}
    \hfill
    \begin{subfigure}[b]{0.3\textwidth}
        \centering
        \includegraphics[width=\textwidth]{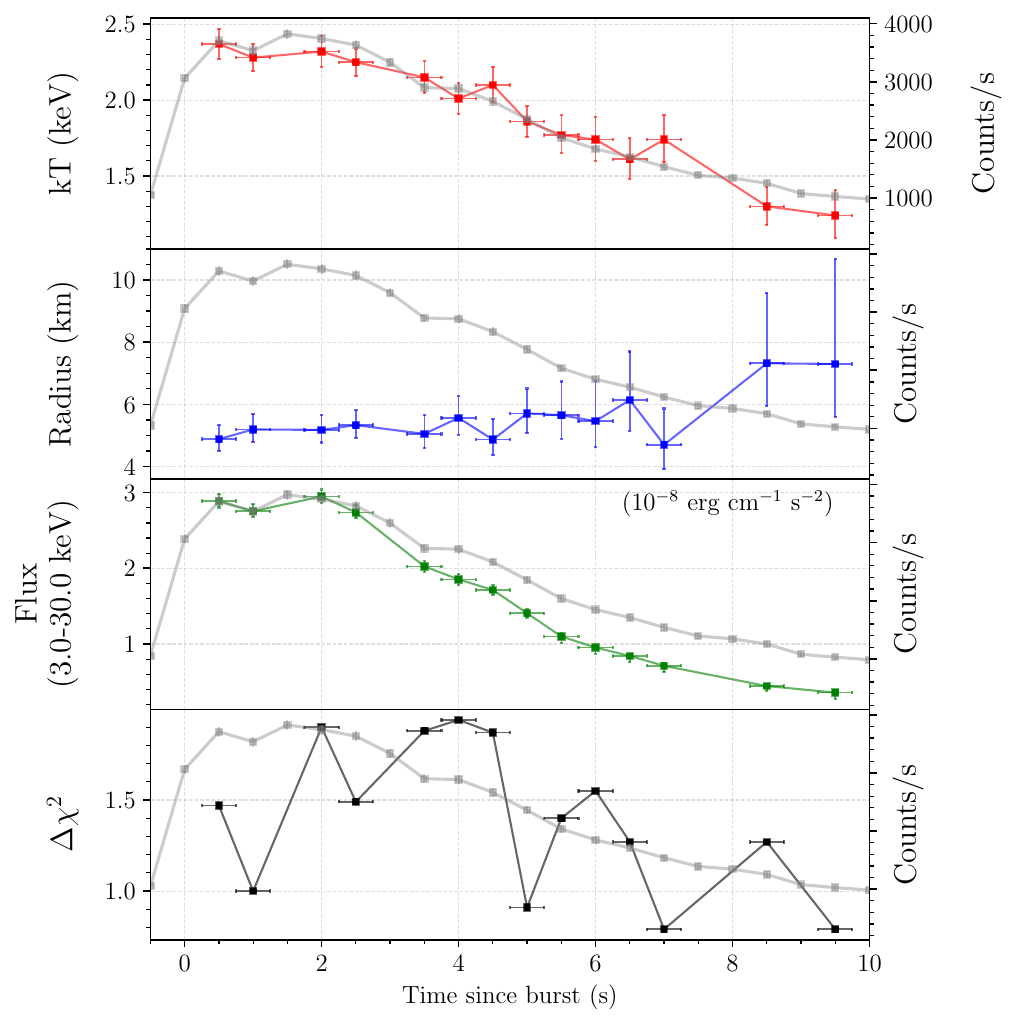}
        \caption{Same as Figure \ref{fig:time_resolved}, for B9}
        \label{fig:3134_burst_1}
    \end{subfigure}
    \hfill
    \begin{subfigure}[b]{0.3\textwidth}
        \centering
        \includegraphics[width=\textwidth]{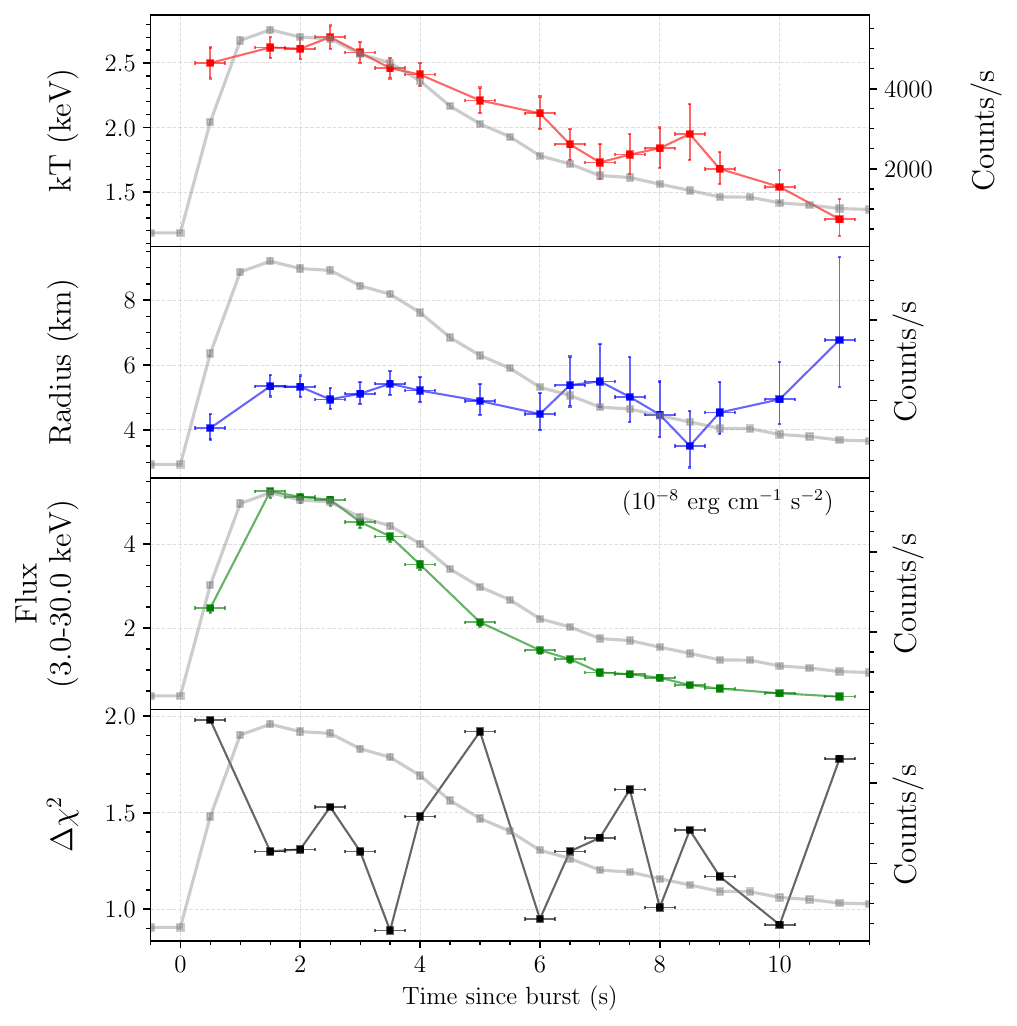}
        \caption{Same as Figure \ref{fig:time_resolved}, for B10}
        \label{fig:3134_burst_2}
    \end{subfigure}
    
    \begin{subfigure}[b]{0.3\textwidth}
        \centering
        \includegraphics[width=\textwidth]{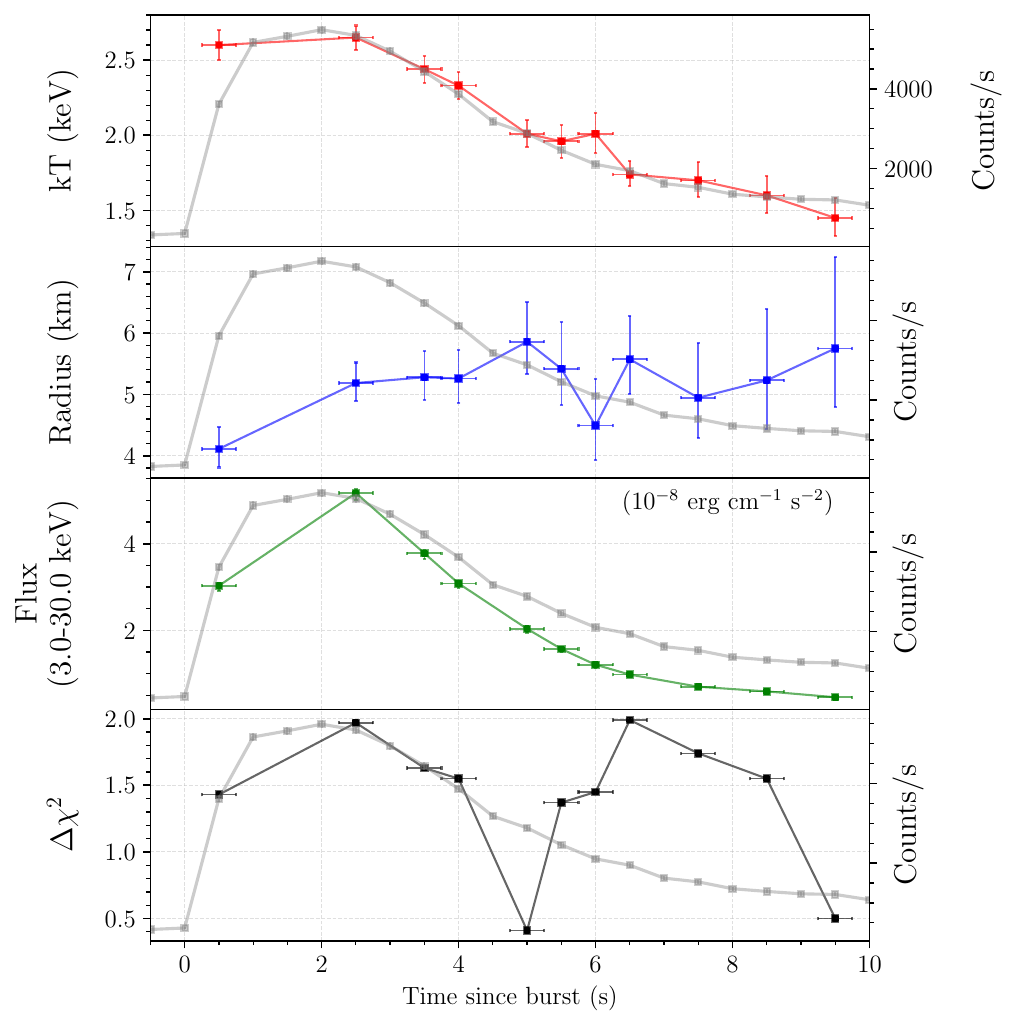}
        \caption{Same as Figure \ref{fig:time_resolved}, for B11}
        \label{fig:3134_burst_3}
    \end{subfigure}
    \hfill
    \begin{subfigure}[b]{0.3\textwidth}
        \centering
        \includegraphics[width=\textwidth]{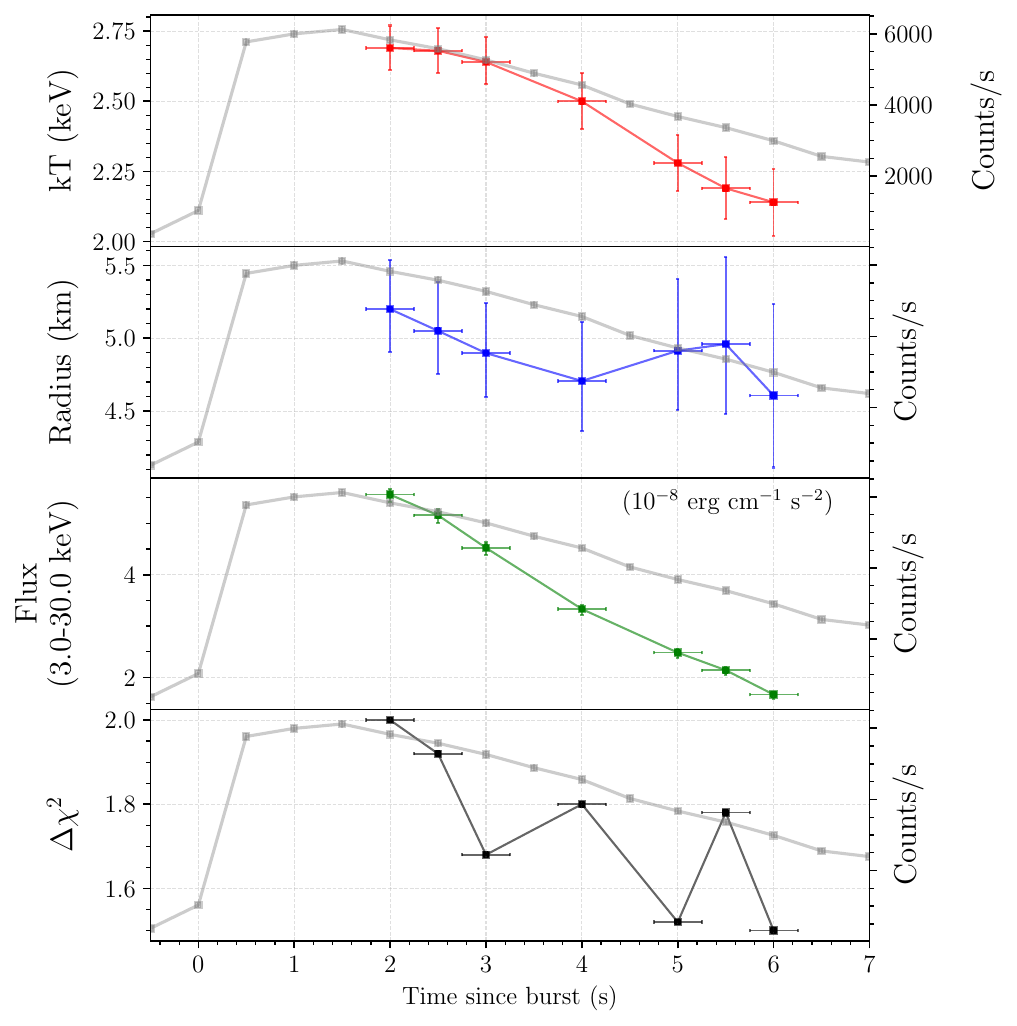}
        \caption{Same as Figure \ref{fig:time_resolved}, for B12}
        \label{fig:3134_burst_4}
    \end{subfigure}
    \hfill
    \begin{subfigure}[b]{0.3\textwidth}
        \centering
        \includegraphics[width=\textwidth]{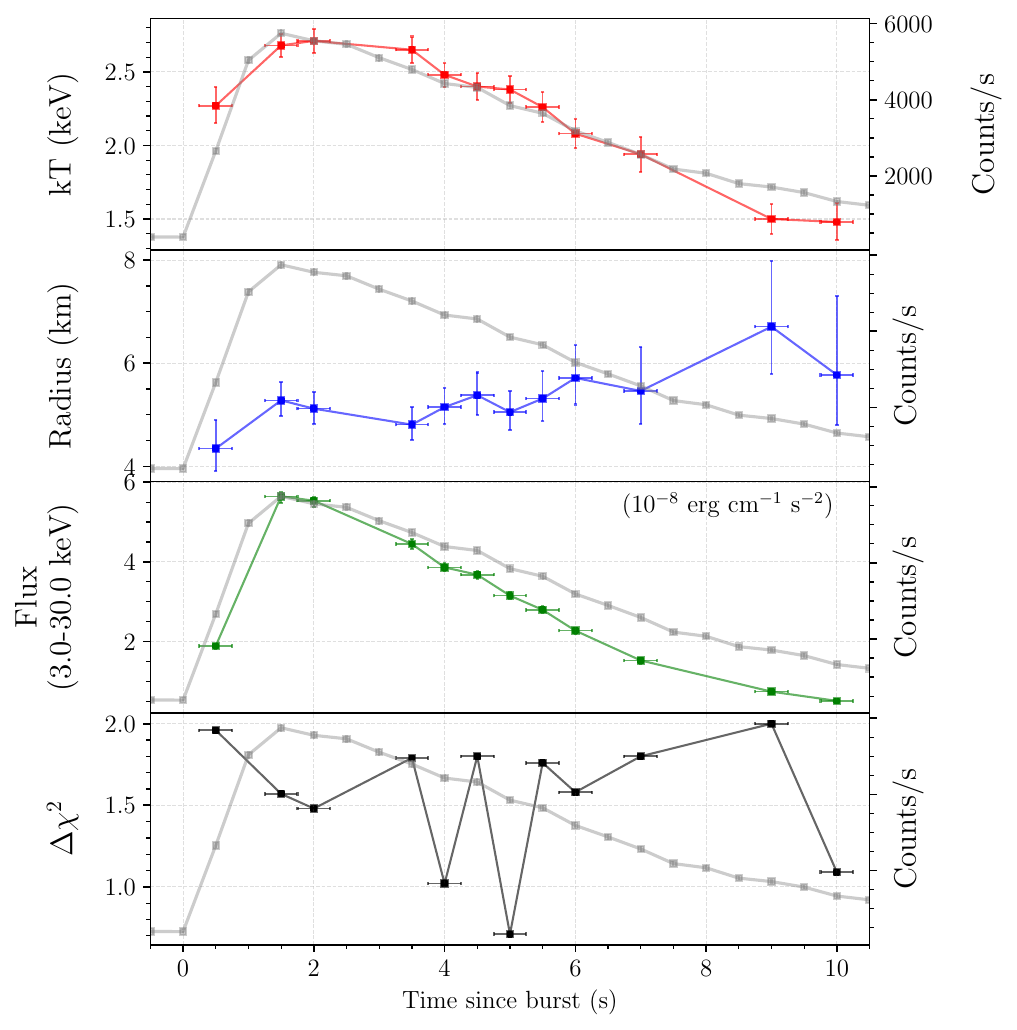}
        \caption{Same as Figure \ref{fig:time_resolved}, for B13}
        \label{fig:3134_burst_5}
    \end{subfigure}
    \vspace{-10pt}
    \caption{Set of 12 burst parameter images (B3).}
    \label{fig:burst_images}
\end{figure*}

\end{document}